\documentclass[aps,onecolumn,prd]{revtex4}
\tighten

 \usepackage{epsf}
\usepackage{color}

\input epsf
\usepackage{graphicx}
\newcommand{\beq}{\begin{equation}}
\newcommand{\beqa}{\begin{eqnarray}}
		  \newcommand{\eeq}{\end{equation}}
\newcommand{\eeqa}{\end{eqnarray}}

\newcommand{\lsim}{\lesssim}
\newcommand{\gsim}{\gtrsim}

\newcommand{\vect}[1]{\mbox{\boldmath${#1}$}}
\newcommand{\lmk}{\left(}
\newcommand{\rmk}{\right)}
\newcommand{\lnk}{\left\{ }
\newcommand{\rnk}{\right\} }
\newcommand{\lkk}{\left[}
\newcommand{\rkk}{\right]}
\newcommand{\lla}{\left\langle}
\newcommand{\p}{\partial}
\newcommand{\rra}{\right\rangle}

\newcommand{\vex}{{\vect x}}

\newcommand{\ven}{\vect n}

\newcommand{\cD}{{\cal D}}
\newcommand{\cM}{{\cal M}}
\newcommand{\hk}{{\hat k}}

\begin{document}

\title{Geometrical Aspects on Parameter estimation of stochastic gravitational wave background: beyond the Fisher analysis} 
%
%
\author{Naoki Seto}
\affiliation{Department of Physics, Kyoto University
Kyoto 606-8502, Japan
}
\author{Koutarou Kyutoku}
\affiliation{Theory Center, Institute of Particles and Nuclear Studies, KEK,
Tsukuba, Ibaraki, 305-0801, Japan
}
\date{\today}

%
%
%
%
\begin{abstract} 
The maximum likelihood method is often used for parameter estimation in 
gravitational wave astronomy. Recently, an interesting  approach was  proposed by 
Vallisneri to evaluate the distributions of parameter estimation errors expected 
for the method. This approach is to statistically analyze  the local peaks of the 
likelihood surface, and  works efficiently even for signals  with low signal-to-noise ratios.
Focusing special attention to geometric structure of the likelihood surface, we follow the 
proposed approach and derive formulae for a simplified model of data analysis where the target 
signal has only one intrinsic parameter, along with its overall amplitude. Then 
we apply our formulae to correlation analysis of stochastic gravitational wave 
background with a power-law spectrum. We report qualitative trends of the 
formulae using numerical results specifically obtained for correlation analysis 
with two Advanced-LIGO detectors.

\end{abstract}
\maketitle

\section{Introduction}

Nowadays, large-scale ground-based laser interferometers such as LIGO
\cite{Harry:2010zz}, Virgo \cite{Accadia:2011zzc} and
KAGRA (formerly LCGT) \cite{Kuroda:2011zz}, are being upgraded or
constructed to realize powerful second generation detectors. It is
expected that we will succeed to directly detect gravitational waves
(GWs) around 10-1000Hz in this decade.  Subsequently, the Laser
Interferometer Space Antenna (LISA) \cite{lisa}
(see also \cite{AmaroSeoane:2012km} for eLISA/NGO) will explore a new
window of GWs around 0.1-100mHz. At the lower frequency regime $\sim
1$nHz, the pulsar timing arrays \cite{Manchester:2011ec,Hobbs:2009yy}
have been significantly improving their sensitivities to GWs.

Under these circumstances, possibilities of GW astronomy have been
actively discussed for these projects, and extracting
parameters characterizing GWs is widely recognized as one of the most
important tasks.  To evaluate the accuracy of parameter estimation, the
Fisher matrix approximation is a standard tool and often used in these
studies \cite{thorne,Finn:1992wt,Cutler:1994ys,jk}. This method is quite
simple to implement, but its performance is known to become worse at
lower signal-to-noise ratios (SNRs) \cite{Vallisneri:2007ev}.
Unfortunately, a full numerical study mimicking actual data analysis
requires a huge computational cost. To fill the gaps between these two
methods, Vallisneri \cite{Vallisneri:2011ts} recently proposed an
interesting and efficient method to predict distributions of the
parameter estimation errors expected for maximum likelihood analyses. He
noticed that the mean densities of the local stationary points (peaks,
valleys and saddle points) of a likelihood surface can be handled
relatively concisely
under fluctuations of the surface induced by the detector
noises. This is because (i) the dependence of the relevant expressions
on the noises is rather simple and (ii) only a small number of
independent noise components is involved. In his work, it was
suggested that the new method can well reproduce the costly results
obtained by fully numerical methods. He also commented that the proposed
method can be utilized to analyze multiple local peaks, including not
global ones that could cause troubles at the actual
parameter estimation.  In this paper, we examine this direction, paying
attention to geometrical properties of likelihood surfaces, not only
their local peaks but also valleys and saddle points.

As a first step, our target is limited to a simple model where we
estimate only one intrinsic parameter and the overall amplitude of the
signal (thus at most two fitting parameters).  While we cannot analyze
important issues inherent to large dimensionalities of fitting
parameters, our study would elucidate
 basic aspects of parameter estimation with the maximum
likelihood method.

In this paper, we first present  a
formal analysis to write down the expected densities of local stationary
points of a likelihood surface. Here we assume Gaussian noises, but do
not specifically limit our analysis to GW observation. Then we apply our
formal results to correlation analysis of stochastic GW background. We
assume a power-law spectrum for the background and discuss estimation of
the spectral index and the overall amplitude.  Many theoretical models of the background
predict power-law spectra, reflecting cosmological or astrophysical
scale-free processes relevant for generation of GWs, and
therefore the assumptions on the spectral shape would
be reasonable at least in the frequency band of a detector (see {\it
e.g.}
\cite{Maggiore:1999vm,Phinney:2001di,Kuroyanagi:2008ye,Nakayama:2008ip,Alabidi:2012ex}). Therefore,
the spectral index and the amplitude would be the primary parameters of
a background and serve as the key information to discriminate its
origin.  Since the SNR of the correlation analysis increases with the
observation time $T_{\mathrm{obs}}$ as $SNR\propto \sqrt{T_{\mathrm{obs}}}$
\cite{Flanagan:1993ix,Allen:1997ad}, we initially need to deal with a
low SNR data. This fact may reduce the validity of the
Fisher matrix analysis for the early era of GW astronomy. Given these
aspects, our simple analysis by the new method with one
intrinsic parameter is not just a toy model, but firmly has a suitable
and realistic application.

As a concrete model,  we examine the correlation analysis with the two Advanced LIGO 
detectors and evaluate the expected number densities of the local 
stationary points of the maximum likelihood surface in our parameter 
space. These results would be useful to discuss the prospects of
stochastic GW background measurements 
with LIGO, and also helps us to grasp qualitative trends of the formal 
expressions. 

{ We find that, for moderate signal strength $SNR\gsim 5$, there
would be vanishingly low probabilities to have multiples peaks on the
likelihood surfaces around the true parameters of the GW background. 
In contrast, false peaks arise  mainly by
noises at the distant parameter regions where the true signal loses
correlation. They typically have low likelihood values and will be
safely excluded by setting an appropriate threshold on the likelihood
value.  We also discuss biases of the fitting parameters estimated with
the maximum likelihood method.  For $SNR\to \infty$, the biases
asymptotically decrease as $1/SNR^2$ relative to the true parameter and
would be buried beneath the parameter estimation errors
($\propto1/SNR$).  }

This paper is organized as follows; in \S II we briefly discuss parameter 
estimation with the maximum likelihood method. In \S III, we provide formal 
expressions for densities of the local stationary points.  \S IV is devoted to 
link the results in \S III to the correlation analysis for stochastic GW 
background. In \S V, we evaluate the densities of the stationary points for the  
two Advanced LIGO detectors and report 
 the observed trends. We also compare the traditional Fisher matrix approximation 
 with the new predictions. \S VI is a summary of this paper.

\section{parameter estimation}

In this section, we briefly discuss a simplified model of 
 data analysis, particularly estimation of characteristic parameters
contaminated by instrumental noise. Our data are given by a real vector
$\mu = ( \mu_1 , \dots , \mu_M )$ with its dimension $M$, and each
element $\mu_\alpha$ ($\alpha$ runs from 1 to $M$) consists of the mean
value $u_\alpha$ and the noise $\nu_\alpha$ as
\begin{equation}
 \mu_\alpha = u_\alpha + \nu_\alpha .
\end{equation}
Throughout this paper, the noise $\nu_\alpha$ is presumed to have a
Gaussian distribution with zero mean and variance $\sigma_\alpha^2$. It
is also assumed that each pair of the noise components has no
correlation, i.e.,
\begin{equation}
 \langle \nu_\alpha \nu_\beta \rangle = \delta_{\alpha \beta} \sigma_\alpha^2 ,
\end{equation}
where the bracket $\langle \rangle$ means the ensemble average.

We define the inner product between two real vectors $a = ( a_1 , \cdots
, a_M )$ and $b = ( b_1 , \cdots , b_M )$ with their dimension $M$ by
\begin{equation}
 \{ a , b \} \equiv \sum_{\alpha=1}^M \frac{a_\alpha
  b_\alpha}{\sigma_\alpha^2}
  . \label{product}
\end{equation}
The probability distribution function for the noise $\nu$ is expressed
using this inner product as
\begin{equation}
 P (\nu) \mathcal{D} \nu = \mathcal{N} \exp \left[ - \frac{\{ \nu , \nu
					     \}}{2} \right] \mathcal{D}
 \nu , \label{pro}
\end{equation}
where $\mathcal{D} \nu = \prod_{\alpha=1}^M d\nu_\alpha$ and
$\mathcal{N} = \prod_{\alpha=1}^M (2\pi
\sigma_\alpha^2)^{-1/2}$. Hereafter, we omit the subscript $\alpha$ of
the vector component for simplicity, whenever we expect that the
confusion of the vector component and the vector itself may not arise.

In this study, candidates of our target signal $u$ are assumed to have
the form
\begin{equation}
 u = \rho \hat{k} (p) , \label{tmpl}
\end{equation}
where $\rho \ge 0$ is the overall amplitude and $\hat{k} (p)$ is the
template vector characterized by a single intrinsic parameter, $p$. The
template is chosen to be a unit vector, so that it satisfies the
normalization condition
\begin{equation}
 \{ \hat{k} (p) , \hat{k} (p) \} = 1 . \label{norm}
\end{equation}
According to the definition described above, the amplitude parameter
$\rho$ is identical to the optimal SNR of the data $\mu$, and has a
clear meaning. In particular, we assign $\rho_{\mathrm{t}}$ and $p_{\mathrm{t}}$ ({t}:
suffix for the true value) for the parameters of the true signal $u_{\mathrm{t}}$
as
\begin{equation}
 u_{\mathrm{t}} = \rho_{\mathrm{t}} \hat{k} ( p_{\mathrm{t}} ) = \rho_{\mathrm{t}} \hat{k}_{\mathrm{t}} ,
\end{equation}
where $\hat{k}_{\mathrm{t}} \equiv \hat{k} (p_{\mathrm{t}})$. While we basically consider the
case in which $\rho \ge 0$, such as a positive-definite power spectrum
in \S IV, we will also provide relevant expressions for general cases
with unconstrained signature of $\rho$, which may be useful for the
analysis of more general aspects, such as the gravitational-wave
polarization.

Our primary task in the data analysis is to estimate the true parameter
$( \rho_{\mathrm{t}} , p_{\mathrm{t}} )$ of the target signal from the contaminated data,
\begin{equation}
 \mu = u_{\mathrm{t}} + \nu ,
\end{equation}
which we can observe in reality. A standard and efficient prescription
is the likelihood analysis, in which template families are prepared to
fit the data. In this study, the template is given by $\rho \hat{k} (p)$
with two parameters $( \rho , p )$, and we define the inner product
\begin{equation}
 \mathcal{M}_{II} ( \rho , p \, ; \nu ) \equiv - \left\{ \mu - \rho
						  \hat{k} (p) , \mu -
						  \rho \hat{k} (p)
						 \right\} ,
\end{equation}
which is closely related to the distance \footnote{The distance should
be defined by $\sqrt{ - \mathcal{M}_{II} ( \rho , p \, ; \nu )}$.}
between the data $\mu$ and the template $\rho \hat{k} (p)$. For a given
noise vector $\nu$, we regard $\mathcal{M}_{II}$ as a continuous
function on the two dimensional plane $( \rho , p )$, and search the
point $( \rho , p ) = ( \rho_{\mathrm{bf}} , p_{\mathrm{bf}} )$ where
the function $\mathcal{M}_{II}$ takes the globally maximum value in the
data analysis. Here, the subscript ``bf'' stands for ``best fit.'' 

The inner product $\mathcal{M}_{II}$ is a quadratic function of the
amplitude $\rho$, and can be written as
\begin{equation}
 \mathcal{M}_{II} ( \rho , p \, ; \nu ) = - \left( \rho - \left\{ \mu ,
							   \hat{k}
							   (p)\right \}
							  \right)^2 +
 \left\{ \mu , \hat{k} (p) \right\}^2 - \left\{ \mu , \mu \right\}
 . \label{dec}
\end{equation}
The first term is the only term dependent on $\rho$, and we can always
set this term to zero by appropriately choosing $\rho$. Therefore, we
initially search the index $p = p_{\mathrm{bf}}$ where the inner product
$\mathcal{M}_I ( p \, ; \nu ) \equiv \{ \mu , \hat{k} (p) \}$ takes its
global maximum \footnote{Actually, the global maximum of the function
$\mathcal{M}_{II}$ is at the parameter $p$ with maximum $|\mathcal{M}_I
( p \, ; \nu )|$. But our concrete model for GW backgrounds analyzed in
this paper has a physical requirement $\rho \ge 0$ (as already assumed).
Therefore, we mostly analyze the simple form $\mathcal{M}_I ( p \, ; \nu
)$ instead of $|\mathcal{M}_I ( p \, ; \nu )|$. But we briefly revisit
this issue in \S III.}, and assign the best-fit amplitude as
\begin{equation}
 \rho_{\mathrm{bf}} = \left\{ \mu , \hat{k} ( p_{\mathrm{bf}} ) \right\} = \mathcal{M}_I (
  p_{\mathrm{bf}} \, ; \nu ) . \label{mi}
\end{equation}
This procedure is essentially the same as the matched filtering analysis
with the normalized templates $\hat{k} (p)$ and the Wiener filter
$\left\{ \mu , \hat{k} (p) \right\}$ (see {\it e.g.}
\cite{thorne}). The simple relation Eq.~(\ref{mi}) between the amplitude
$\rho_{\mathrm{bf}}$ and the peak value $\mathcal{M}_I ( p_{\mathrm{bf}} )$ turns out to
be useful later. Hereafter, we omit the argument $\nu$ of
$\mathcal{M}_{II}$ and $\mathcal{M}_I$ for simplicity. {  The subscripts ``$I$'' 
and ``$II$'' represent the dimensions of the fitting  parameters (``$I$'' 
for the single parameter $p$ and ``$II$'' for the two parameters $(\rho,p)$). }

The estimated values $(\rho_{\mathrm{bf}},p_{\mathrm{bf}})$ depend on specific realization
of the noise $\nu$, and are scattered around the true values
$(\rho_{\mathrm{t}},p_{\mathrm{t}})$. Therefore, they should be regarded as statistical
variables fluctuating in response to the realizations of the noise
vector $\nu$. Our primary interest in this paper is the probability
distribution function of the estimated parameters $(\rho_{\mathrm{bf}},p_{\mathrm{bf}})$.

At the global solution $(\rho,p)=(\rho_{\mathrm{bf}},p_{\mathrm{bf}})$ obtained for a
given noise vector $\nu$, the function $\mathcal{M}_I$ meets the
following relations required for a {\it local} peak,
\begin{equation}
 \partial_p \left\{ \mu,\hat{k} (p) \right\} = 0 , ~~
  \partial_p^2\left\{ \mu , \hat{k} (p) \right\} < 0 , \label{peak}
  \end{equation}
as necessary conditions \footnote{
The simple expressions in this paper are given for data analysis with a
single intrinsic parameter $p$.  If there are totally $N_p$ intrinsic
parameters $p_1 , p_2 , \cdots , p_{N_p}$, the local peaks of the function
$\{ \hat{k} , \mu \}$ are the stationary points $\partial_{p_i} \{
\hat{k} , \mu \} = 0$ where all the eigenvalues of the $N_p\times N_p$
Hesse matrix $\partial_{p_i} \partial_{p_j} \{ \hat{k} , \mu \}$ are
negative.}. However, the local relations Eq.~(\ref{peak})
are not the sufficient conditions for the {\it global} maximum of the
function $\mathcal{M}_I (p)$, as it might have multiple peaks for a
single realization of the noise $\nu$.  With multiple peaks, it is
necessary to select the global maximum in actual data analysis.

{ Nevertheless, it was shown in \cite{Vallisneri:2011ts} (see  
Fig.3 in the paper)  
that  numerical results for distribution of the global peaks of likelihood 
surfaces can be reproduced well by a local expression that actually counts the stationary points of the surfaces.} Based 
on this 
observation, the aims of this paper are (i) to geometrically develop an
analytical framework for the local peak statistics in simplified
one-dimensional cases, and (ii) to apply it for the correlation analysis
of GW backgrounds, as a realistic example.

In our local approach, we unavoidably count the contribution of more
than one peaks of the function $\mathcal{M}_I (p)$. In general, it is
difficult to analytically handle global properties of complicated
functions (see {\it e.g.} \cite{adler}).  On the other hand, between two
adjacent peaks of a one-dimensional function, we must have a valley
(local minimum) with the relations
\begin{equation}
 \partial_p \left\{ \mu , \hat{k} (p) \right\} = 0 , ~~ \partial_p^2
  \left\{ \mu , \hat{k} (p) \right\} > 0 , \label{val} 
\end{equation}
because of the continuity of the function $\partial_p \left\{ \mu ,
\hat{k} (p) \right\}$. These two are local conditions, and can be
managed analytically. We thus analyze the distribution of the valleys
that would supplementary help us to discuss the multiplicity of the
solutions $p$ for the local peaks Eq.~(\ref{peak}).

Next, based on the above discussions on the peaks and valleys of the
one-dimensional function $\mathcal{M}_I (p)$, we expand our
considerations to the local geometry on the two-dimensional surface
$\mathcal{M}_{II} (\rho,p)$. Here, it should be noted that the cross
section of the surface $\mathcal{M}_{II}(\rho,p)$ at a fixed parameter
$p$ has a parabolic shape convex upward with $\partial^2
\mathcal{M}_{II} / \partial \rho^2 = -2 < 0$. Therefore, no local
minimum on the two dimensional surface $\mathcal{M}_{II} ( \rho , p )$
appears. Indeed, the parabolic shape along the amplitude $\rho$ is the
universal feature of any dimensional likelihood surface as long as
normalized template families are adopted.

For a solution $p=p_{\mathrm{pk}}$ of the local peak conditions Eq.~(\ref{peak}),
we assign the corresponding amplitude by $\rho_{\mathrm{pk}} = \mathcal{M}_I
(p_{\mathrm{pk}})$. Then, the function $\mathcal{M}_{II} (\rho,p)$ turns out to
have a local peak at $(\rho_{\mathrm{pk}},p_{\mathrm{pk}})$ as easily seen from
Eq.~(\ref{dec}). In the same manner, we can assign the amplitude
$\rho_{\mathrm{vl}} = \mathcal{M}_I (p_{\mathrm{vl}})$ for a solution $p=p_{\mathrm{vl}}$ of the
local valley conditions Eq.~(\ref{val}). Although the function
$\mathcal{M}_{II}(\rho,p)$ becomes a saddle point (not a local minimum)
at the point $(\rho_{\mathrm{vl}},p_{\mathrm{vl}})$, we continue to use the suffix ``vl''
originally defined for the valleys of the one-dimensional function
$\mathcal{M}_I (p)$ in this two-dimensional case.

Although we only deal with the real data $\mu$ in this paper, it is
straightforward to expand our formalism for complex data with random
Gaussian noises. For complex vectors $a$ and $b$, the inner product
Eq.~(\ref{product}) should be modified as
\begin{equation}
 \{ a , b \} = \frac{1}{2} \sum_\alpha \frac{a_\alpha b_\alpha^* +
  a_\alpha^* b_\alpha}{\sigma_\alpha^2} ,
\end{equation}
and the elements $\mathcal{D} \nu$ and $\mathcal{N}$ should be modified
to include both real and imaginary contributions of the noise,
$\nu$. The amplitude $\rho$ should also be regarded as a complex
variable, and we can still make similar arguments for parameter
estimation based on the relation
\begin{equation}
 \mathcal{M}_{II} = - \left| \rho - \left\{ \mu , \hat{k} (p) \right\}
		      \right|^2 + \left| \left\{ \mu , \hat{k} (p)
					 \right\} \right|^2 - \{ \mu ,
		      \mu \} .
\end{equation}

\section{Densities of local peaks}

As commented earlier, our data $\mu=\rho_{\mathrm{t}} \hk(p_{\mathrm{t}})+\nu$ contain the
 noise $\nu$ that results in fluctuating the positions of the local
 peaks.  Now, let us consider an ensemble of the noise vectors $\nu$
 whose probability distribution function is given by Eq.(\ref{pro}). For
 each realization of the noise vector $\nu$, we can
 pick up all the local peaks for the fluctuated function
 $\cM_I(p)$. Here the total number of the local peaks is not necessarily
 unity.  Next, for the ensemble of the noises vectors, we statistically
 handle the spatial distributions of the local peaks. In this manner we
 can evaluate the expected number of the local peaks in a small
 parameter range $[p,p+\delta p]$ and express it in the form \beq
 \sigma_{\mathrm{pk}}(p) \delta p.  \eeq Due to its definition, we can regard
 $\sigma_{\mathrm{pk}}(p)$ as the expected number density of the local peaks.

Similarly, we put the expected number of the local peaks for the function 
$\cM_{II}(\rho,p)$ in a two dimensional region $[\rho,\rho+\delta \rho]\times 
[p,p+\delta p]$ by
\beq
\sigma_{\mathrm{pk}}(\rho,p) \delta \rho~ \delta p
\eeq
with the corresponding number density $\sigma_{\mathrm{pk}}(\rho,p)$.
In this section, basically following \cite{Vallisneri:2011ts}, we derive analytical 
 expressions $\sigma_{\mathrm{pk}}(p)$ as well as $\sigma_{\mathrm{pk}}(\rho,p)$ for the 
 expectation values of the local peaks.  
 Considering potential multiplicity of the local peaks, we call
 these functions as densities, rather than the probabilities (that should be 
 normalized to unity).

Here it is important note that (i)  the global 
 peaks are  sub-classes of the local peaks and (ii) our density distributions 
 $\sigma_{\mathrm{pk}}(p)$ and  $\sigma_{\mathrm{pk}}(\rho,p)$ 
 would provide upper limits for the probability distributions of the global ones.
In the same manner, we denote  the expected number densities of 
 local valleys (and saddles) by  $\sigma_{\mathrm{vl}}(p)$ and   $\sigma_{\mathrm{vl}}(\rho,p)$.

In this section, we do not use the concrete form of the normalized template 
$\hk(p)$. Therefore, our results in this section can be generally applicable for 
estimation of a single parameter $p$ and  the associated amplitude 
$\rho$, through the relation (\ref{dec}) under presence of Gaussian noises.

\subsection{formal expressions}

First, we introduce the simplified notations $k^{(i)}(p)$  ($i=0,1,2$) below 
\beq
k^{(i)}(p)\equiv \p_p^i \hk(p)
\eeq
for the derivatives of the unit template
vector $\hk(p)$ with
$k^{(0)}\equiv\hk$ for $i=0$.

For a given noise vector $\nu$, we can count the number ${\cal N}(p \,; \nu)\delta 
p$ of the local  
peaks  in the parameter range $[p,p+\delta p]$ for  the function 
$\cM_I(p)=\{\hk(p) , \mu\}$ as (see {\it e.g.} \cite{Vallisneri:2011ts,adler,Bardeen:1985tr})
\beq
{\cal N}(p \,; \nu)\delta p=\int_p^{p+\delta p} dp~ \delta_{\mathrm D}\lkk \p_p 
\cM_I(p\, ;\nu)\rkk
T\lkk -\p_p^2 \cM_I(p \, ; \nu) \rkk \label{cn}
\eeq
where $\delta_{\mathrm D}$ is the delta function and we defined the function
\beq
T(x)=\cases{ 
    0 & ($x\le0$) \cr
    x  & ($x>0$) \cr
}.
\eeq
In Eq.(\ref{cn}), the delta function represents the condition for the extremum 
$\p_p\cM_I=0$, and we temporarily recover the argument $\nu$ for the function 
$\cM_I$ in order to clarify its 
dependence on the noise.  The function 
$T$ selects the sign $\p_p^2 \cM<0$  appropriate for a peak, and also  fixes the measure 
associated with the delta function.
Taking account of the probability distribution of the noise $\nu$, the expected 
number of the local peaks is given by 
\beq
\sigma_{\mathrm{pk}}(p)\delta p=\int \cD \nu P(\nu) {\cal N}(p;\nu)\delta p=\delta p \int \cD\nu P(\nu)  \delta_{\mathrm D}\lkk \p_p \cM_I(p)\rkk
T\lkk -\p_p^2 \cM_I(p) \rkk,
\eeq
or equivalently
\beq
\sigma_{\mathrm{pk}}(p)=\int \cD\nu P(\nu) \delta_{\mathrm D}\lkk \lnk k^{(1)}(p),\mu \rnk\rkk
T\lkk -\lnk k^{(2)}(p),\mu \rnk \rkk \label{2pk}.
\eeq

  In the same  manner, the density of the local
valleys is given by 
\beq
\sigma_{\mathrm{vl}}(p)=\int \cD\nu  P(\nu)\delta_{\mathrm D}\lkk \lnk k^{(1)}(p),\mu \rnk\rkk
T\lkk \lnk k^{(2)}(p),\mu \rnk \rkk. \label{2vl}
\eeq

As for the two dimensional density distribution of the local peaks and saddles 
(with the subscript ``vl''), we have similar expressions
\beq
\sigma_{\mathrm{pk}}(\rho,p)=\int \cD \nu P(\nu)  \delta_{\mathrm D}\lmk\rho-\lnk \hk,\mu \rnk\rmk 
\delta_{\mathrm D}\lkk \lnk k^{(1)},\mu \rnk\rkk
T\lkk -\lnk k^{(2)},\mu \rnk \rkk \label{3pk}
\eeq
and
\beq
\sigma_{\mathrm{vl}}(\rho,p)=\int \cD \nu P(\nu)  \delta_{\mathrm D}\lmk\rho-\lnk \hk,\mu \rnk\rmk 
\delta_{\mathrm D}\lkk \lnk k^{(1)},\mu \rnk\rkk
T\lkk \lnk k^{(2)},\mu \rnk \rkk. \label{3vl}
\eeq

The above expressions (\ref{2pk})(\ref{2vl})(\ref{3pk}) and (\ref{3vl}) are written as multidimensional integrals $\cD \nu$ for 
 the noise vector $\nu$.  However,  for a given parameter $p$, only the following three inner products $
N_0\equiv \lnk \hk(p),\nu\rnk ,~N_1\equiv \lnk k^{(1)}(p),\nu\rnk $ and 
$N_2\equiv \lnk k^{(2)}(p),\nu\rnk 
$  are relevant in Eqs.(\ref{3pk}) and (\ref{3vl}).
For Eqs.(\ref{2pk}) and (\ref{2vl}), we need to deal with only the 
two combinations $N_1$ and $N_2$.  

The variables $N_0$, $N_1$ and $N_2$ are specific linear combinations of the 
large-dimensional vector $\nu$.
Therefore, the actual dimensions of the 
integral $\cD \nu$ can be reduced down to 3 or 2 \cite{Vallisneri:2011ts}.  If the each component $\nu_\alpha$ of 
the  noise vector is Gaussian, the probability distribution function $P(N_0,N_1,N_2)$ is completely determined by their covariance 
matrix $\lla N_i N_j \rra$.  From the definition of the inner product, we have
\beq
\lla N_i N_j\rra=\lla \{k^{(i)}(p),\nu\} \{k^{(j)}(p),\nu\}  \rra=C_{ij}(p),
\eeq
where we defined
\beq
C_{ij}=C_{ji}\equiv \lnk k^{(i)}(p), k^{(j)}(p)  \rnk. \label{cij}
\eeq
From the normalization  $\{\hk,\hk\}=1$ of the templates, we readily have
$C_{10}=0$ and $C_{11}+C_{20}=0$.  We also define the product $D_i(p)$ between the vector 
$k^{(i)}(p)$ and the unit vector $\hk_{\mathrm{t}}\equiv \hk(p_{\mathrm{t}})$ for the true index $p_{\mathrm{t}}$ as
\beq
D_i(p)\equiv \lnk k^{(i)}, \hk_{\mathrm t}  \rnk=\p_p^i D_0(p).\label{di}
\eeq

Integrating out irrelevant noise elements in Eq.(\ref{2pk}),  the density 
$\sigma_{\mathrm{pk}}(p)$ is given by 
\beq
\sigma_{\mathrm{pk}}(p)=\int dN_1 dN_2 P(N_1,N_2) \delta_{\mathrm D}[N_1+\rho_{\mathrm{t}} D_1] T[-\rho_{\mathrm{t}} D_2-N_2].
\eeq
While we can directly manage this expression, the covariance $C_{21}\ne 0$ 
between $N_1$ and $N_2$ is somewhat cumbersome for polynomial deformations 
\footnote{We can use the functional 
freedom of the parameter $p$  to 
simplify the covariance matrix for the noises. 
{ More specifically, we introduce the new parameter $q$ with the relation
$
{dq}/{dp}=\sqrt{C_{11}(q)}.
$
Then we have $C_{00}'=C_{11}'=1$ and $C_{01}'=C_{21}'=0$.
Here the quantities with the prime $'$ are given for the new parameter $q$.
The only non-trivial one  $C_{22}'$ is written with the original ones $C_{ij}$ 
(for the parameter $p$) by
$
C_{22}'=({C_{22}C_{11}-C_{21}^2})/{C_{11}^3}$. 
We can easily deal with the probability distribution function of the related 
noise matrix due to the simple structure of the correlation  $C_{ij}'$ without 
using  
the additional vector $N_{\mathrm{oth}}$.  Once we derive the density
 $\sigma_{\mathrm{pk}}'(q)$ for the new variable $q$. The density for  the 
 original parameter $p$ is 
given by $\sigma_{\mathrm{pk}}(p)=\sigma_{\mathrm{pk}}'(q)dq/dp$.}}. 
  Below, we take a different route by 
introducing the new unit vector $\hk_{\mathrm {oth}}$ defined by
\beq
\hk_{\mathrm {oth}}(p)\equiv\frac{C_{11}k^{(2)}-C_{21}k^{(1)}}{\sqrt{C_{11}(C_{22}C_{11}-C_{21}^2)}}
\eeq
that satisfies $\lnk \hk_{\mathrm {oth}},\hk_{\mathrm {oth}}  \rnk=1$ and is orthogonal to the vector $k^{(1)}$ as 
$\lnk \hk_{\mathrm {oth}},k^{(1)} \rnk=0$.

The original vector $k^{(2)}$ is  given by $k^{(1)}$ and $\hk_{\mathrm {oth}}$ as 
\beq
k^{(2)}(p)=\frac{\sqrt{C_{11}(C_{22}C_{11}-C_{21}^2)}\hk_{\mathrm {oth}}+C_{21}k^{(1)} }{C_{11}}.
\eeq
We hereafter use $\hk_{\mathrm {oth}}$ instead of $k^{(2)}$, and define the products $X_i$ and 
$Y$ by
\beq
X_i(p)\equiv \{k^{(i)},\hk_{\mathrm {oth}}\},~~Y(p)\equiv \{\hk_{\mathrm{t}},\hk_{\mathrm {oth}}\}.
\eeq
They are given  by the products $C_{ij}$ and $D_i$ as
\beq
X_0=-\frac{C_{11}^2}{\sqrt{C_{11} (C_{11} C_{22}-C_{21}^2)}},~~X_1=0,~~X_2=\sqrt{\frac{ C_{11} C_{22}-C_{21}^2}{C_{11}}}\label{xi}
\eeq
and
\beq
Y=\frac{C_{11}D_2-C_{21}D_1}{\sqrt{C_{11}(C_{22}C_{11}-C_{21}^2)}}=\frac{D_2}{X_
2}-\frac{C_{21} D_1}{X_2 C_{11}}.\label{y}\label{y}
\eeq
We also introduce the new stochastic variable $N_{\mathrm {oth}}$ as 
\beq
N_{\mathrm {oth}}=\lnk \hk_{\mathrm {oth}},\nu  \rnk =\frac{N_2}{X_2}-\frac{C_{21}N_1}{X_2 C_{11}}.\label{not}
\eeq
We have $\lla N_{\mathrm {oth}}, N_1 \rra=\{\hk_{\mathrm {oth}},k^{(1)}\}=0$ and $\lla N_{\mathrm {oth}}, N_0\rra=\{\hk_{\mathrm {oth}},k^{(0)}\}=X_0$. Then the covariance matrix 
between $(N_0,N_1,N_{\mathrm {oth}})$ is given by 
\beq
F= \pmatrix{
\lla N_0 N_0\rra  & \lla N_0 N_1\rra & \lla N_0 N_{\mathrm {oth}}\rra   \cr
\lla N_1 N_0\rra  & \lla N_1 N_1\rra & \lla N_1 N_{\mathrm {oth}}\rra   \cr
\lla N_{\mathrm {oth}} N_0\rra  & \lla N_{\mathrm {oth}} N_1\rra & \lla N_{\mathrm {oth}} N_{\mathrm {oth}}\rra  \cr 
}
= \pmatrix{
1 & 0 &X_0 \cr
0 & C_{11} & 0 \cr
X_0 & 0 & 1  \label{matr}
}.
\eeq
Taking inverse of the relevant parts of the matrix, we have the probability 
distribution functions as
\beq
P(N_1,N_{\mathrm {oth}})=\frac1{2\pi \sqrt{C_{11}}}\exp\lmk -\frac{N_1^2}{2C_{11}} \rmk \exp\lmk -\frac{N_{\mathrm {oth}}^2}{2} \rmk 
\eeq
and 
\beq
P(N_0,N_1,N_{\mathrm {oth}})=\frac1{(2\pi)^{3/2} \sqrt{C_{11}(1-X_0^2)}}\exp\lmk 
-\frac{N_{\mathrm {oth}}^2+N_0^2-2X_0 N_0 N_{\mathrm {oth}}}{2(1-X_0^2)} \rmk \exp\lmk -\frac{N_1^2}{2C_{11}} \rmk. 
\eeq

From the formal expression (\ref{2pk}), 
we eliminate the variables $N_2$ and $D_2$ using Eqs.(\ref{y}) and (\ref{not}), and obtain
\beq
\sigma_{\mathrm{pk}}(p)=\int dN_1 dN_{\mathrm {oth}} P(N_1,N_{\mathrm {oth}}) 
\delta_{\mathrm D}[\rho_{\mathrm{t}} D_1+N_1] T[-X_2(\rho_{\mathrm{t}} 
Y+N_{\mathrm {oth}})-C_{21}(\rho_{\mathrm{t}} D_1+N_{1})/C_{11}]. \label{fepk}
\eeq
By performing the $N_1$-integral first, we find 
\beq
\sigma_{\mathrm{pk}}(p)=\frac1{ \sqrt{2\pi C_{11}}} \exp\lmk -\frac{\rho_{\mathrm{t}}^2 D_1^2}{2C_{11}} 
\rmk X_2 F_{\mathrm{pk}}(-\rho_{\mathrm{t}} Y)\label{2pka}
\eeq
with
\beqa
F_{\mathrm{pk}}(a)&\equiv&\int^{\infty}_{-a}dx\frac{(x+a)e^{-\frac{x^2
   }{2}}}{\sqrt{2 \pi
   }}\\
&=&\int^{\infty}_{0}dx\frac{xe^{-\frac{(x-a)^2
   }{2}}}{\sqrt{2 \pi
   }}\\
&=&\frac{a}{2} 
   {\rm erfc}\left(-\frac{a}{\sqrt{2
   }}\right)+\frac{e^{-\frac{a^2
   }{2}}}{\sqrt{2 \pi
   }}. \label{erfc}
\eeqa
Here, the first factor $\exp(-\rho_{\mathrm{t}}^2 D_1^2/2C_{11})$  originates from 
the delta function for a stationary point, and is closely related to 
 the Fisher matrix prediction (see
the next subsection). { In Eq.(\ref{erfc}) we  used the  complementary error 
function ${\rm erfc}(x)\equiv 1-{\rm erf}(x)=2\int_z^\infty e^{-t^2}dt/\sqrt{\pi}$. }
In the same manner we obtain the density of the local valleys as
\beq
\sigma_{\mathrm{vl}}(p)=\frac1{ \sqrt{2\pi C_{11}}} \exp\lmk -\frac{\rho_{\mathrm{t}}^2 D_1^2}{2C_{11}} 
\rmk X_2 F_{\mathrm{vl}}(-\rho_{\mathrm{t}} Y) \label{2vla}
\eeq
with
\beq
F_{\mathrm{vl}}(a)\equiv \int^{-a}_{-\infty}dx\frac{-(x+a)e^{-\frac{x^2
   }{2}}}{\sqrt{2 \pi
   }}.\label{fvl}
\eeq
The two functions $F_{\mathrm{pk}}$ and $F_{\mathrm{vl}}$ are plotted in Fig.1. We can easily 
derive the following relations
\beq
F_{\mathrm{pk}}(a)=F_{\mathrm{vl}}(-a),~~~F_{\mathrm{pk}}(a)-F_{\mathrm{vl}}(a)=a,~~~F_{\mathrm{pk}}(0)=F_{\mathrm{vl}}(0)=\frac1{\sqrt{2 \pi
   }},~~~\lim_{a\to \infty}\frac{F_{\mathrm{pk}}(a)}{a}=1.
\eeq
We have $\sigma_{\mathrm{pk}}/\sigma_{\mathrm{vl}}=F_{\mathrm{pk}}/F_{\mathrm{vl}}$ for the relative abundances of 
the peaks and valleys.  The number of 
peaks dominates that of the valleys at $-
\rho_\mathrm{t} Y > 0$.

The two dimensional density profiles $\sigma_{\mathrm{pk}}(\rho,p)$ and  $\sigma_{\mathrm{vl}}(\rho,p)$  can be evaluated 
similarly as
\beqa
\sigma_{\mathrm{pk}}(\rho,p)&=&\frac{{\sqrt{1-X_0^2}}}{2\pi \sqrt{  C_{11}}} \exp\lmk -\frac{\rho_{\mathrm{t}}^2 D_1^2}{2C_{11}} 
\rmk\exp\lmk -\frac{(\rho-\rho_{\mathrm{t}} D_0)^2}{2} 
\rmk X_2 F_{\mathrm{pk}}\lmk -\frac{\rho_{\mathrm{t}} Y+X_0 (\rho-\rho_{\mathrm{t}} D_0)}{\sqrt{1-X_0^2}}\rmk \label{3pka},\\
\sigma_{\mathrm{vl}}(\rho,p)&=&\frac{{\sqrt{1-X_0^2}}}{2\pi \sqrt{  C_{11}}} \exp\lmk -\frac{\rho_{\mathrm{t}}^2 D_1^2}{2C_{11}} 
\rmk\exp\lmk -\frac{(\rho-\rho_{\mathrm{t}} D_0)^2}{2} 
\rmk X_2 F_{\mathrm{vl}}\lmk -\frac{\rho_{\mathrm{t}} Y+X_0 (\rho-\rho_{\mathrm{t}} D_0)}{\sqrt{1-X_0^2}}\rmk \label{3vla},
\eeqa
and we have the following identity
 between $\sigma_{\mathrm{pk}}(p)$ and $\sigma_{\mathrm{pk}}(\rho,p)$
\beq
\sigma_{\mathrm{pk}}(p)=\int_{-\infty}^{\infty}d\rho ~\sigma_{\mathrm{pk}}(\rho,p) \label{integral}
\eeq
due to the simple structure for  the amplitude parameter $\rho$.

While we have introduced the 
orthogonal vector $\hk_{\mathrm {oth}}$ to simplify the covariance noise matrix (\ref{matr}), we 
can directly reach Eqs.(\ref{2pka})(\ref{2vla}) and (\ref{3pka}) from the 
original expressions (\ref{2pk})(\ref{2vl}) and (\ref{3pk}).

{ We evaluate the mean value of the peak amplitude $\rho$ as follows
\beqa
{\bar \rho}_{\mathrm{pk}}(p)&\equiv &\frac{\int_{-\infty}^\infty \rho ~\sigma_{\mathrm{pk}}(\rho,p) d\rho}{\sigma_{\mathrm{pk}}(p)}\\
&=&\rho_{\mathrm t} D_0-X_0 \frac{1+{\rm erf}(-\rho_{\mathrm 
t}Y/\sqrt{2})}{2 F_{\mathrm{pk}}(-\rho_{\mathrm t}Y)}.\label{pmean}
\eeqa
Here the first term $\rho_{\mathrm t} D_0$ is the simple average of the product 
$\{\hk(p),\mu\}=\rho_{\mathrm t}D_0(p)+N_0$ with respect to the noise $N_0$. The 
second one is a positive definite term  with a negative factor $X_0=O(1)$, and 
represents the bias caused by selecting only peaks. Due to the negative 
correlation $\lla N_0 N_2 \rra=-\lla N_1 N_1 \rra<0$ between the two noise 
components $N_0$ and $N_2$, the requirement for being a peak (related to $N_2$) 
introduces the  bias for the product $\{\hk(p),\mu\}=\rho_{\mathrm t}D_0(p)+N_0$. }

\begin{figure}
  \begin{center}
\epsfxsize=7.5cm
\begin{minipage}{\epsfxsize} \epsffile{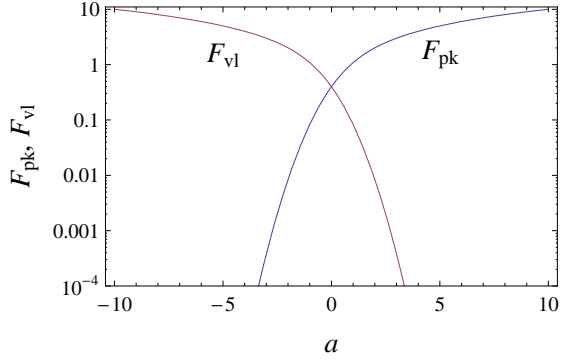}\end{minipage}
 \end{center}
  \caption{The functions $F_{\mathrm{pk}}(a)$ and $F_{\mathrm{vl}}(a)$ for densities of local peaks and valleys.
 }
\end{figure}

 As mentioned earlier, at actual data analysis, we 
initially search the point $p=p_{\mathrm{bf}}$ where the product  $\cM_I=\{\hk(p),\mu\}$ 
takes the global maximum.   Even if our target points are shifted to the local 
peaks of the function $\cM_I(p)$, instead of the global peak,  it is  expected 
that the product $\cM_I(p)$ would take relatively 
large values for local peaks around the true parameter $p_{\mathrm{t}}$ but smaller values for 
those generated merely by statistical fluctuations at points distant from $p_{\mathrm{t}}$.
 In  this manner, the 
magnitude of the product $\cM_{I}(p)$ at a local peak would become in itself an
useful indicator for our theoretical  analysis purely based on local quantities.  Here we 
introduce the notation  
$\cM_{{I \mathrm{pk}}}$ for the value of $\cM_I (p_{\mathrm{pk}})$ at a local peak and distinguish  it from 
the original one-dimensional function $\cM_I(p)$.

We thus consider the expected number of local peaks in the parameter range 
$[p,p+\delta p]$ and the peak height $[\cM_{{I \mathrm{pk}}},\cM_{{I \mathrm{pk}}}+\delta \cM_{I \mathrm{pk}}]$, 
and denote it by $s_{\mathrm{pk}}(\cM_{I \mathrm{pk}},p)\delta p ~\delta \cM_{I \mathrm{pk}}$. 
Following the arguments around Eqs.(\ref{cn})-(\ref{2pk}) we have 
\beq
s_{\mathrm{pk}}(\cM_{I \mathrm{pk}},p)\equiv \int \cD \nu P(\nu)  \delta_{\mathrm D}\lmk\cM_{I \mathrm{pk}}-\lnk \hk,\mu \rnk\rmk 
\delta_{\mathrm D}\lkk \lnk k^{(1)},\mu \rnk\rkk
T\lkk -\lnk k^{(2)},\mu \rnk \rkk \label{3spk}.
\eeq 
But 
 the expression (\ref{3spk}) is essentially the 
same as Eq.(\ref{3pk}) 
\beq
s_{\mathrm{pk}}(\cM_{I \mathrm{pk}},p)=\sigma_{\mathrm{pk}}(\rho=\cM_{I \mathrm{pk}},p)
\eeq
due to the simple correspondence between the estimated amplitude $\rho$ 
and the inner product $\cM_I$ as shown in Eq.(\ref{mi}).  Therefore, we can use the density distribution $\sigma_{\mathrm{pk}}(\rho,p)$
 also for the function $s_{\mathrm{pk}}(\cM_{I \mathrm{pk}},p)$.  
Once the curves $s_{\mathrm{pk}}(\cM_{I \mathrm{pk}},p_1)$ and $s_{\mathrm{pk}}(\cM_{I \mathrm{pk}},p_2)$ are given 
for two different values $p=p_1$ and  
$p_2$, we can apply the relation $\int_{-\infty}^\infty dy 
s(y,p)=\sigma_{\mathrm{pk}}(p)$ (see Eq.(\ref{integral})) to compare the relative densities 
of $\sigma_{\mathrm{pk}}(p_1)$ and $\sigma_{\mathrm{pk}}(p_2)$ by eye, based on the areas of the two curves.

The unimportant peaks due to noises at an index  $p$ distant  from the true value 
$p_{\mathrm{t}}$ would mostly have low peak heights and would  be efficiently 
removed by choosing an appropriate threshold on $\cM_{I \mathrm{pk}}$, as demonstrated later.
To elucidate this, 
we  define the density of local  peaks above a given threshold by
\beq
\sigma_{\mathrm{pk}}(>\cM_{I \mathrm{pk}},p)\equiv \int_{\cM_{I 
\mathrm{pk}}}^\infty dy ~s(y,p)=\int_{\cM_{I \mathrm{pk}}}^\infty 
d\rho~ \sigma_{\mathrm{pk}}(\rho,p) \label{pkth}.
\eeq

As commented earlier, we first searched maximums of the function $\cM_I(p)$ 
instead of $|\cM_I(p)|$, considering the requirement $\rho\ge 0$ valid {\it e.g.} 
 for the estimation of a power spectrum that is a positive definite quantity (as 
 analyzed in the next section). 
  If we literally evaluate 
the local peaks/valleys for the function $\cM_{II}(\rho,p)$ without the prior $\rho>0$,  they are given 
with our expressions (\ref{3pka}) and (\ref{3vla})  as
\beq
\sigma_{\mathrm{pk}}(\rho,p)\theta(\rho)+\sigma_{\mathrm{vl}}(\rho,p)\theta(-\rho),~~\sigma_{\mathrm{pk}}(\rho,p)\theta(-\rho)+\sigma_{\mathrm{vl}}(\rho,p)\theta(\rho)\label{re1}
\eeq
 respectively.  Here $\theta(x)$ is the step function.
 Similarly, the local peaks/valleys of the function $|\cM_I(p)|$ 
 are obtained as
\beq
\int_{-\infty}^\infty d\rho \lkk 
\sigma_{\mathrm{pk}}(\rho,p)\theta(\rho)+\sigma_{\mathrm{vl}}(\rho,p)\theta(-\rho)\rkk,~~\int_{-\infty}^\infty d\rho\lkk \sigma_{\mathrm{pk}}(\rho,p)\theta(-\rho)+\sigma_{\mathrm{vl}}(\rho,p)\theta(\rho)\rkk.\label{re2}
\eeq
Here, we should notice that the roles of peaks and valleys of the 
 function  $\cM_I(p)$ interchange for the absolute value $ |\cM_I(p)|$ at $\cM_I(p)<0$.
While we do not use these somewhat complicated  expressions (\ref{re1}) and (\ref{re2}), these would 
be more adequate, depending on  problems.

\subsection{large SNR limit}

It is well known that, with a large SNR, the distribution of the 
parameters estimated by the matched filtering  is well approximated by the Fisher matrix 
predictions  around their true values \cite{Finn:1992wt,Cutler:1994ys,Vallisneri:2007ev}. In this subsection, we examine the profiles of our density distribution 
functions $\sigma_{\mathrm{pk}}(p)$ and $\sigma_{\mathrm{pk}}(\rho,p)$ at larger $\rho_{\mathrm{t}}$.  
Similar analyses were already done in \cite{Vallisneri:2011ts}, but it would be 
instructive to directly examine our analytic expressions obtained in the 
previous subsection.

We 
first expand the fitting parameter around their true values and define the 
deviations as
\beq
\Delta p\equiv p-p_{\mathrm{t}},~~~\Delta \rho\equiv \rho-\rho_{\mathrm{t}}.
\eeq
Then, taking the leading order term with respect to $\Delta p$, we obtain
\beq
D_0\simeq1,~~D_1\simeq -C_{11{\mathrm{t}}}\Delta p,~~D_2\simeq -C_{11{\mathrm{t}}}
\eeq
and
\beq
Y\simeq \frac{-C_{11{\mathrm{t}}}^2}{\sqrt{C_{11{\mathrm{t}}}(C_{22{\mathrm{t}}}C_{11{\mathrm{t}}}-C_{21{\mathrm{t}}}^2)}}<0
\eeq
where the product $C_{ij{\mathrm{t}}}=\{k^{(i)}(p_{\mathrm{t}}), k^{(j)}(p_{\mathrm{t}})\}$ is evaluated at the point
$p=p_{\mathrm{t}}$. With the asymptotic relation $F_{\mathrm{pk}}(a)\sim a$ at 
$a\to \infty$,  
the expressions (\ref{2pka}) and (\ref{3pka}) for the local peaks can be approximated as
\beq
\sigma_{\mathrm{pk}}(p)\simeq \frac1{{\sqrt {2\pi  C_{11\mathrm{t}}^{-1}\rho_{\mathrm{t}}^{-2}}}} \exp\lkk-\frac{\Delta p^2}{2C_{11\mathrm{t}}^{-1} 
\rho_{\mathrm{t}}^{-2}}  \rkk \label{fish}
\eeq
and
\beq
\sigma_{\mathrm{pk}}(\rho,p)\simeq \frac1{2\pi {\sqrt {C_{11\mathrm{t}}^{-1}\rho_{\mathrm{t}}^{-2}}}} \exp\lkk-\frac{\Delta p^2}{2C_{11\mathrm{t}}^{-1} 
\rho_{\mathrm{t}}^{-2}}  \rkk \exp\lkk -\frac{(\Delta \rho)^2}{2} \rkk \label{fish2}
\eeq
for  small $|\Delta p|$ and $|\Delta \rho|$ and at $\rho_{\mathrm{t}}\gg 1$.

Meanwhile we have the Fisher matrix for the  two parameters $\rho$ and $p$ at 
their true values as
\beq
\pmatrix{
\lnk \p_\rho (\rho \hk),\p_\rho (\rho \hk)  \rnk  &  \lnk \p_\rho (\rho \hk),\p_p (\rho \hk)  \rnk   \cr
\lnk \p_p (\rho \hk),\p_\rho (\rho \hk)  \rnk & \lnk \p_p (\rho \hk),\p_p (\rho \hk)  \rnk  \cr
}_{\rho_{\mathrm{t}},p_{\mathrm{t}}}
=\pmatrix{ 1 & 0 \cr
     0& \rho_{\mathrm{t}}^2C_{11{\mathrm{t}}}    \cr
}.
\eeq
It is straightforward to confirm that the Fisher matrix predictions
agree with our expressions 
(\ref{fish}) and (\ref{fish2}) originally given for the local peaks. We hereafter 
denote the 
right-hand sides of these equations by $\sigma_{\mathrm{fisher}}(p)$ and $\sigma_{\mathrm{fisher}}(\rho,p)$.

At $\rho_{\mathrm{t}} \to \infty$, the Gaussian  distribution  $\sigma_{\mathrm{fisher}}(p)$ is strongly localized around $\Delta p=0$ with 
the characteristic width $\propto \rho_{\mathrm{t}}^{-1}$.  Therefore it would be 
advantageous to use the rescaled variable $x\equiv \rho_{\mathrm{t}} \Delta p$ to analyze 
the shape of the function $\sigma_{\mathrm{pk}}(p)$ relative to  
$\sigma_{\mathrm{fisher}}(p)$.   After some algebra, we can derive the following 
perturbative expression.
\beq
\sigma_{\mathrm{pk}}(p)=\sigma_{\mathrm{fisher}}(p)+\rho_{\mathrm{t}}^0\eta(x)+{\rm H.O.}.
\eeq
Here we have $\sigma_{\mathrm{fisher}}(p)\propto \rho_{\mathrm{t}}$ and the higher order term H.O. 
is given by a polynomial of $x$ whose coefficients are at most $O(\rho_{\mathrm{t}}^{-1})$.
The leading-order correction term $\eta(x)$ is given by 
\beq
\eta(x)=\frac{C_{21{\mathrm{t}}}}{\sqrt{2\pi 
C_{11{\mathrm{t}}}}}\exp\lkk -\frac{C_{11{\mathrm{t}}}x^2}2  
\rkk\lkk x-\frac{C_{11{\mathrm{t}}}}2x^3\rkk \label{eta}.
\eeq
  Thus, with 
the rescaled variable $x$, the difference $\sigma_{\mathrm{pk}}(p)-\sigma_{\mathrm{fisher}}(p)$ 
asymptotically approaches   the fixed function $\eta(x)$ at $\rho_{\mathrm{t}}\to 
\infty$.  In the next section, we  demonstrate this numerically.

In order to characterize the shape of the function  $\sigma_{\mathrm{pk}}(p)$, we evaluate 
its zeroth, first and second moments  by  taking
integrals. { 
Since $\eta(x)$ is an odd function, we can derive the following results; 
\beqa
\int \sigma_{\mathrm{pk}}(p)~ d\Delta p&=&1+O(\rho_{\mathrm t}^{-2}),\label{mean0}\\
\int \sigma_{\mathrm{pk}}(p) \Delta p~ d\Delta p&=&-\frac12\rho_{\mathrm{t}}^{-2} C_{21{\mathrm{t}}}C_{11{\mathrm{t}}}^{-2}+O(\rho_{\mathrm{t}}^{-3}),\label{mean}\\
\int \sigma_{\mathrm{pk}}(p) (\Delta p)^2~d\Delta p&=&C_{11{\mathrm{t}}}^{-1}\rho_{\mathrm{t}}^{-2}+O(\rho_{\mathrm{t}}^{-4}).\label{vari}
\eeqa
These would be used in \S V.D.  Due to the normalization condition (\ref{mean0}), 
the right-hand-side of Eq.(\ref{mean}) can be regarded as the estimation bias of 
the primary parameter $p$.
In the same manner, we have the bias for the mean value for the overall amplitude
\beq
\int \sigma_{\mathrm{pk}}(p) [{\bar \rho}_{\mathrm{pk}}(p)-\rho_{\mathrm t} ] d\Delta p=\frac1{2\rho_{\mathrm{t}}}+O(\rho_{\mathrm{t}}^{-2}).\label{meana}
\eeq
Note that the term $1/({2\rho_{\mathrm{t}}})$ is  a second order correction $O(\rho_{\mathrm{t}}^{-2})$ relative to the true amplitude $\rho_{\mathrm{t}}$.
The parameters estimated by the maximum likelihood method generally have biases from the 
second order $O(\rho_{\mathrm{t}}^{-2})$ (see \cite{Vitale:2010mr} for a 
perturbative analysis) and those in Eqs.(\ref{mean}) and (\ref{meana}) agree with 
results obtained from perturbative expressions for the effects of the noises 
({\it e.g.} Eq.(A31) in \cite{Cutler:1994ys}).
}

\section{correlation analysis for stochastic GW backgrounds}
Hereafter, we apply our formal studies to correlation 
analysis of gravitational wave background \cite{Flanagan:1993ix,Allen:1997ad}. In this section, we first describe basic aspects of the
correlation analysis, and mention its correspondence to the data analysis 
prescription discussed in \S II and III. Then we provide expressions that would be useful for 
numerically evaluating the local peak/valley densities for parameter estimation 
of GW backgrounds with power-law spectra.

\subsection{data correlation}

We discuss observation of an isotropic stochastic GW background with two L-shaped 
detectors $I$ and $J$  
in an observational period $T_{\mathrm{obs}}$. 
The Fourier modes of two data streams  $s_{I,J}(f)$ are 
linear combinations of the responses to the background signal  $h_{I,J}(f)$  and 
the detector 
noises  $n_{I,J}(f)$ as
\beq
s_I(f)=h_I(f)+n_I(f),~~s_J(f)=h_J(f)+n_J(f).
\eeq
We assume that the detector noises $n_I(f)$ and $n_J(f)$ are stationary with no 
correlation between them  (namely $\lla n_I(f)^* n_J(f') \rra=0$). We define  the noise  
spectra of the two detectors 
$P_I(f)$ and $P_J(f)$  in the following relations
\beq
\lla n_I(f)^* n_I(f') \rra=\frac12  P_I(f)\delta_{\mathrm D}(f-f'),~~
\lla n_J(f)^* n_J(f') \rra=\frac12  P_J(f)\delta_{\mathrm D}(f-f').
\eeq
 The responses  
$h_I(f)$ and  $h_J(f)$ to the GW 
background  would have correlation that is characterized by the overlap 
reduction function $\gamma_{IJ}(f)$  as 
\beq
\lla h_I(f)^*h_J(f') \rra=\frac{3H_0^2 \gamma_{IJ}(f)}{20\pi^2 f^3} 
\Omega_{\mathrm{GW}}(f)\delta_{\mathrm D}(f-f')
,\label{corh}
\eeq 
where $\Omega_{\mathrm{GW}}(f)$ is the energy density of the GW background in the 
logarithmic frequency interval and normalized by the critical density of the 
universe $3H_0^2/8\pi$ ($H_0$: the Hubble parameter hereafter fixed at 
70km/sec/Mpc). 
The overlap reduction function $\gamma_{IJ}$ depends strongly on the relative 
configuration of the two detectors and is given by the following 
angular integral \cite{Flanagan:1993ix,Allen:1997ad}
\beq
\gamma_{IJ}(f)=\frac5{8\pi}\int_{S^2}d\ven [F_I^{+*}F_J^+ 
+F_I^{\times*}F_J^\times] \exp[2\pi i f (\vex_J-\vex_I)\cdot \ven]\label{orf}
\eeq
with the beam pattern functions $F_{I,J}^{+,\times}$ and the spatial positions of 
detectors $\vex_{I,J}$. We have the upper limit $|\gamma_{IJ}|=1$  valid for 
co-aligned detectors.

In this 
article we  study the situations where the observational data $s_I(f)$ is dominated by the detector 
noises ($|h_I|\ll |n_I|$) as
\beq
\frac{3H_0^2\Omega_{\mathrm{GW}}(f)}{10\pi^2f^3}\ll P_I(f),~P_J(f)
\eeq
 (weak signal 
condition).   Under this condition, the correlation analysis becomes an efficient approach to examine a weak 
GW background. 

Following the prescription described in \cite{Seto:2005qy},
we divide the observational frequency band into finite segments $F_{\alpha}$ 
( $\alpha=1,\cdots,{L}$: the suffix for the segments) that 
have the widths $\delta f_{\alpha}$ and the central frequencies $f_{\alpha}$. 
The widths $\delta f_{\alpha}$ are selected to satisfy $T_{\mathrm{obs}}^{-1}\ll \delta 
f_\alpha\ll f_\alpha$ so that, in  each segment, (i) there are a large number 
($\delta f_\alpha/T_{\mathrm{obs}}^{-1}\gg 1$)
of Fourier modes, and (ii) the frequency dependencies can be neglected ($\delta 
f_\alpha/f_\alpha\ll 1$) for the 
functions, such as $P_I(f)$, 
$P_J(f)$, $\gamma_{IJ}(f)$ and $\Omega_{\mathrm{GW}}(f)$.  These two conditions  hold for 
the laser interferometers such as LIGO \cite{Harry:2010zz}, Virgo 
\cite{Accadia:2011zzc}, KAGRA \cite{Kuroda:2011zz}, BBO \cite{Corbin:2005ny,Harry:2006fi} 
and DECIGO \cite{Seto:2001qf,Kawamura:2006up}, but not 
for the  pulsar timing experiments which are sensitive  at $f\sim T_{\mathrm{obs}}^{-1}$.
Meanwhile, we have $\gamma_{IJ}(f)=0$ for independent data streams of LISA \cite{Prince:2002hp,Krolak:2004xp}.

To statistically amplify the target background signals and compress the data, we take the summation 
of the data products in each segment $\alpha$ as
\beq
\mu_{\alpha}={\rm Re}\lkk \sum_{f\in F_{\alpha}} s_I(f)^* s_J(f)\rkk.\label{mu}
\eeq
Here we decompose $\mu_\alpha$ in terms of its mean 
value $u_{\alpha}$ and  statistical
fluctuation with zero mean $\nu_{\alpha}$ as
\beq
\mu_{\alpha}=u_{\alpha}+\nu_{\alpha}.
\eeq
At this stage, we do not need to be aware of the relation between notations 
introduced here and in \S II and III.
  From Eq.(\ref{corh}) the mean $u_\alpha$ is given by   
\beq
u_{\alpha}\equiv \lla \mu_{\alpha} \rra= \lla\sum_{f\in F_{\alpha}} s_I(f)^* 
s_J(f)  \rra=\frac{3H_0^2 \gamma_{IJ}(f_\alpha)\Omega_{\mathrm{GW}} (f_{\alpha})}{20\pi^2 f_\alpha^3}
 \frac{\delta 
f_{\alpha}}{T_{\mathrm{obs}}^{-1}} \label{amu}.
\eeq
Note that the mean value of the summation  $ \sum_{f\in F_{\alpha}} s_I(f)^* 
s_J(f)$  is a real number even without the operator ${\rm Re}[\cdot]$ in  
Eq.(\ref{mu}). This is the reason why we took the real part of the product in  
Eq.(\ref{mu}) to dispose the irrelevant imaginary part of the fluctuation $\nu_\alpha$.  With the  weak signal condition, the fluctuation $\nu_{\alpha}$ is dominated 
by the detector noises and its variance is given by 
\beq
\sigma_{\alpha}^2=\lla \nu_{\alpha}^2 \rra\simeq \lla \lmk {\rm Re}\lkk \sum_{f\in 
F_{\alpha}}n_I(f)^*n_J(f)\rkk \rmk^2 \rra=P_I(f_{\alpha})P_J(f_{\alpha})\frac{\delta f_{\alpha}}{8T_{\mathrm{obs}}^{-1}}.
\eeq
In the last expression, we had an additional factor 1/2 associated with the 
operator ${\rm Re}[\cdot]$ in Eq.(\ref{mu}). The product ${\rm Re}\lkk n_I(f)^*n_J(f) 
\rkk$ at a single frequency $f$ would not be Gaussian distributed. However,  due 
to a large number of involved modes $\delta f_\alpha/T_{\mathrm{obs}}^{-1}\gg 1$ in a segment and 
 the central limit theorem, the fluctuations $\nu_\alpha$ for the compressed data  can be regarded as Gaussian.

Given the noise level $\sigma_\alpha$,  we can evaluate the  SNR of each segment  as
\beq
SNR_{\alpha}^2=\frac{u_{\alpha}^2}{\sigma_{\alpha}^2}=\lkk \frac{3H_0^2 \gamma_{IJ}(f_\alpha)}{10\pi^2 f_\alpha^3} \rkk^2 T_{\mathrm{obs}} \frac{2 \delta f_{\alpha} 
\Omega_{\mathrm{GW}}(f_{\alpha})^2}{P_I(f_{\alpha}) P_J(f_{\alpha})}.\label{snrp}
\eeq
Then the total SNR is given by a quadratic  summation of all the segments 
\beq
SNR^2=\sum_{\alpha=1}^{L} SNR^2_{\alpha}=\lmk \frac{3H_0^2 }{10\pi^2} \rmk^2 T_{\mathrm{obs}} \lkk 
 2 \int_0^\infty df \frac{\gamma_{IJ}(f)^2 \Omega_{\mathrm{GW}}(f)^2}{f^6 
 P_I(f)P_J(f)}\rkk. \label{snrt}
\eeq
Note that the final expression (\ref{snrt}) does not depend on the details of the segmentation, 
and agrees with those in the literature \cite{Flanagan:1993ix,Allen:1997ad}. The total SNR in Eq.(\ref{snrt}) is also expressed as
\beq
SNR^2=\lnk u,u \rnk
\eeq
with the product defined in Eq.(\ref{product}).  Now the vectors $\mu,u,\nu$ 
introduced in this section can be directly  regarded as those in \S 
II and III.  Here, the dimension $M$ of the vectors is  the number 
of the segments $L$.


In this paper, as  concrete models,  we only deal with  the background spectra $\Omega_{\mathrm{GW}}(f)$ 
given in  a power-law form
\beq
\Omega_{\mathrm{GW}}(f)\propto f^p
\eeq
in the frequency band observed by the detectors in interest. Here $p$ is the spectral 
index in the band, and serves as the single intrinsic parameter in the previous sections. 
 For the mean value $u_\alpha$ of the correlation analysis (see 
Eq.(\ref{amu})), we define the unit vector $\hk(p)$ whose components (including 
sign information) 
are given as
\beq
\hk_\alpha(p)\propto  \frac{3H_0^2 \gamma_{IJ}(f_\alpha)}{20\pi^2 f_\alpha^3} {f_\alpha}^p 
\frac{\delta f_\alpha}{T_{\mathrm{obs}}^{-1}} \label{ukv}
\eeq
with the normalization condition
$\lnk \hk(p),\hk(p)\rnk=1$. 
Introducing the additional parameter $\rho(\ge 0)$ for the amplitude of a GW 
background, we express the  
mean value of the correlated data due to the background as $u=\rho \hk(p)$
 \footnote{While we have the physical requirement 
$\rho\ge 0$, we do not impose the corresponding (and also other) priors 
$\rho_{\mathrm{bf}}\ge 0$ for simplicity.}.  We can now apply the formal expressions 
derived in \S III.

In this paper, we only study the weak signal case with two available detectors.  But the 
expression (\ref{snrt}) 
can be extended for stronger GW backgrounds by the following replacement \cite{Allen:1997ad}
\beq
 P_I(f) P_J(f)\to  P_I(f) P_J(f)+\frac{3H_0^2 }{10\pi^2 }  \frac{\Omega_{\mathrm{GW}}}{f^3} 
 (P_I+P_J)+\lmk  \frac{3H_0^2 }{10\pi^2 } \rmk^2 \frac{\Omega_{\mathrm{GW}}^2}{f^6} (1+\gamma_{IJ}^{2}).
\eeq
For correlation analysis with more than two independent detectors, the optimal 
SNR is given by a summation of Eq.(\ref{snrt}) with respect to all the 
possible pairs of detectors. 

Hereafter,  for notational simplicity, we omit the subscripts $I$ and $J$ for our two
detectors, and use expressions, such as
$\gamma=\gamma_{IJ}$.

\subsection{shape function}

As shown in Eqs.(\ref{2pka})(\ref{2vla}) and (\ref{3pka}), our expressions for 
the local peaks and valleys are given by the inner products $C_{ij}$ and $D_i$ 
(see Eqs.(\ref{cij}) and (\ref{di}) for 
their definitions). Here, note that, with Eqs.(\ref{xi}) and (\ref{y}), the 
parameters $X_i$ and $Y$ are 
written in terms of $C_{ij}$ and $D_i$.  In this section, we  provide simple formulae 
that would be easily applicable when numerically evaluating the basic 
ingredients  $C_{ij}$ and $D_i$ for the power law spectra $\Omega_{\mathrm{GW}}\propto f^p$, 
as  in the next section.

Since  $C_{ij}$ and $D_i$ are defined by the inner products of two unit vectors 
and their derivatives with respect to the spectral indexes (see Eqs.(\ref{cij}) 
(\ref{di}) and (\ref{ukv})), they should be 
generated from the  integrals
\beq
\int_{0}^\infty df \frac{\gamma(f)^2 f^{x}}{f^6 P_I(f)P_J(f)}
\eeq
and its (up to the fourth) derivatives with the parameter $x$. Therefore, we define 
the following five functions 
($i=0,1,2,3,4$)
\beq
w_{i}(x)\equiv  A\int_0^\infty df \frac{\gamma(f) (f/f_{\mathrm c})^{x}(\ln [f/f_{\mathrm c}])^i}{f^6 P_I(f)P_J(f)}.
\eeq
Here the frequency $f_{\mathrm c}$ is a characteristic frequency in the observational band 
in interest, and should be set arbitrarily.
For convenience at later discussions, we fix the normalization factor $A$ by the condition
\beq
w_0(0)=1
\eeq
or equivalently put
\beq
w_{i}(x)=\frac{\int_{0}^\infty df \frac{\gamma(f)^2 (f/f_{\mathrm c})^{x}(\ln [f/f_{\mathrm c}])^i}{f^6 P_I(f)P_J(f)}}{\int_{0}^\infty df \frac{\gamma(f)^2 (f/f_{\mathrm c})^{0}}{f^6 P_I(f)P_J(f)}}.\label{wi}
\eeq
  Once 
the lowest-order function $w_0(x)$  is given numerically ({\it e.g.} with a fitting formula 
as in the next section),  we can generate other ones (for $i=1,\cdots,4$) by taking derivatives as
\beq
w_{i}(x)=\p_x^i w_{0}(x).
\eeq
We call $w_0(x)$ as the shape function. This function depends on the profile 
 of the noise spectra $P_{I,J}(f)$, the overlap reduction function $\gamma(f)$, and the selected 
 frequency $f_{\mathrm c}$.   Under the simple geometrical representation for 
 the signal vector $\rho \hk(p)$ with the amplitude  parameter $\rho$,
 most of the principal
 information relevant  for our analyses is included in the shape function.

Now we express the product $C_{ij}=\{k^{(i)}(p),k^{(j)}(p)\}$  in  
terms of $w_{i}(x)$.  Here we need to call the functions $w_{i}(x)$ only at $x=2p$, and 
 define 
\beq
m_i\equiv w_{i}(2p)
\eeq
to simplify our expressions.  
After some algebra,  we can derive
\beq
C_{11}=\frac{m_2 m_0-m_1^2}{m_0^2},~~
C_{21}=\frac{2 m_1^3-3 m_2 m_1
   m_0+m_3m_0^2 }{m_0^3},
~~C_{22}=\frac{-3 m_1^4+6 m_2 m_1^2 m_0-4
   m_3 m_1 m_0^2 +
   m_4m_0^3}{m_0^4}.
\eeq
These combinations do not depend on $A$ and $f_{\mathrm c}$, as 
expected from the simple geometric meanings of the normal vectors.

In order to evaluate $D_i$ related to the true vector $\hk_{\mathrm{t}}=\hk(p_{\mathrm{t}})$, we similarly define the elements $m_{0\mathrm{t}}$ and 
$l_i$ by
\beq
m_{0\mathrm{t}}\equiv w_0(2p_{\mathrm{t}}),~~~~
l_i\equiv w_i (p+p_{\mathrm{t}}).
\eeq
Then we have
\beq
D_0=\frac{l_0}{(m_{0\mathrm{t}}m_0)^{1/2}},~~
D_1=\frac{-m_1 l_0+m_0 l_1}{(m_{0\mathrm{t}}m_0^3)^{1/2}},~~
D_2=\frac{3 m_1^2 l_0-2 m_1 m_0 l_1 +m_0 (m_0 l_2-2 m_2 l_0)}{(m_{0\mathrm{t}}m_0^5)^{1/2}}.
\eeq

So far, we have used the parameter $\rho$ to represent the amplitude of the 
background. This is a geometrically natural choice.  But, in some cases, it might be 
  preferable to put the background spectrum in the form 
\beq
\Omega_{\mathrm{GW}}(f)=\omega_{\mathrm{GW}} \lmk \frac{f}{f_{\mathrm c}} \rmk^p,
\eeq
and 
use the combination of the parameters $(\omega_{\mathrm{GW}},p)$ for discussing prospects 
 of correlation analysis, 
  instead of the original one $(\rho,p)$. 
Below, we summarize expressions related to these two parameterizations.  From Eq.(\ref{snrt}), the two 
  amplitudes $\rho$ and $\omega_{\mathrm{GW}}$ are related by
\beq
\rho^2=\lmk \frac{3H_0^2 }{10\pi^2} \rmk^2 T_{\mathrm{obs}} \omega_{\mathrm{GW}}^2\lkk 
 2 \int_0^\infty df \frac{\gamma(f)^2  \lmk \frac{f}{f_{\mathrm c}} \rmk^{2p}}{f^6 P_I(f)P_J(f)}\rkk.
\eeq

Since the shape function $w_0(x)$ is normalized as
\beq
w_{0}(x)=\frac{\int_{0}^\infty df \frac{\gamma(f)^2 (f/f_{\mathrm c})^{x}}{f^6 P_I(f)P_J(f)}}{\int_{0}^\infty df \frac{\gamma(f)^2 (f/f_{\mathrm c})^{0}}{f^6 P_I(f)P_J(f)}},
\eeq
we have
\beq
\rho= B \omega_{\mathrm{GW}} T_{\mathrm{obs}}^{1/2} [w_0(2p)]^{1/2}
\eeq
with a constant factor $B$ that is determined 
by the noise spectra and the overlap reduction function  as
\beq
B\equiv \lkk 2 \lmk \frac{3H_0^2 }{10\pi^2} \rmk^2 
\int_{0}^\infty df \frac{\gamma(f)^2}{f^6 P_I(f)P_J(f)}\rkk^{1/2}.\label{facb}
\eeq

From the basic property of the delta function, the density 
  $\sigma_{\mathrm{pk}}'(\omega_{\mathrm{GW}},p)$ of the local peaks in the parameter space $(\omega_{\mathrm{GW}},p)$ is expressed 
  with the density $\sigma_{\mathrm{pk}}(\rho,p)$ defined for the original space $(\rho,p)$ as
\beqa
\sigma_{\mathrm{pk}}'(\omega_{\mathrm{GW}},p)&=&\int_{-\infty}^\infty d\rho 
\sigma_{\mathrm{pk}}(\rho,p)\delta_{\mathrm D}\lmk \omega_{\mathrm{GW}}-\frac{\rho}{  B[T_{\mathrm{obs}}w_0(2p)]^{1/2} }\rmk\\
&=&B  T_{\mathrm{obs}}^{1/2} [w_0(2p)]^{1/2} \sigma_{\mathrm{pk}}\lmk B \omega_{\mathrm{GW}} T_{\mathrm{obs}}^{1/2} [w_0(2p)]^{1/2},p\rmk.
\eeqa

\section{correlation analysis with the Advanced LIGO }

In this section, we evaluate our analytical expressions for the two 4km 
Advanced-LIGO detectors, as a concrete example of correlation analysis for 
stochastic gravitational wave backgrounds.  
Throughout this section, we set the characteristic frequency $f_{\mathrm c}$ at
\beq
f_{\mathrm c}=25{\rm Hz}
\eeq
for our power-law spectrum 
$\Omega_{\mathrm{GW}}(f)=\omega_{\mathrm{GW}} \lmk \frac{f}{f_{\mathrm c}} \rmk^p$.

\subsection{basic quantities}
 Our aim in this subsection is to 
 provide  the shape function $w_0(x)$ and related quantities for the two Advanced 
 LIGO detectors.  These are the
  preliminary calculations used for our main results given in the following subsections.

First, 
 the overlap 
 reduction  
function $\gamma(f)$ in Eq.(\ref{orf}) is expressed analytically as  
\cite{Flanagan:1993ix,Allen:1997ad} (see also \cite{Seto:2008sr,Nishizawa:2009bf} 
for polarized modes)
\beq
\gamma(f)= \Theta_1(y,\beta)\cos(4\delta)+
\Theta_2(y,\beta) \cos(4\Delta). \label{gi}
\eeq
The parameter $\beta$ is the angle between the two detectors measured from the 
center of the Earth, and $(\delta,\Delta)$ characterizes the orientations of the 
detectors relative to the great circle connecting the two sites.  The variable 
$y$ is given by 
\beq
y\equiv\frac{2\pi f D}{c}
\eeq
 with the distance $D=2R_E \sin(\beta/2)$  ($R_E=6400$km: the radius of the Earth). The two functions $\Theta_1$ and $\Theta_2$ are written as
\beq
 \Theta_1(y,\beta)=\cos^4\lmk\frac{\beta}2  \rmk \lmk j_0+\frac57
j_2+\frac{3}{112} j_4 \rmk 
\eeq
\beq
 \Theta_2(y,\beta)=\lmk -\frac38 j_0+\frac{45}{56}
j_2-\frac{169}{896} j_4 \rmk
+\lmk \frac12 j_0-\frac57j_2-\frac{27}{224}j_4 \rmk \cos\beta
+ \lmk-\frac18 j_0-\frac5{56}j_2-\frac{3}{896}j_4  \rmk \cos(2\beta)
\eeq
with the spherical Bessel functions $j_n=j_n(y)$. We have the upper limit 
$|\gamma|\le 1$ and  
the equality here holds only for two co-aligned 
detectors (mod $\pi/2$) at a same place ($|\cos 4\delta|=1$ and $\beta=0$).

 For 
the two LIGO detectors, the angular parameters are $\beta=27.2^\circ$,  
$\delta=45.3^\circ$ and $\Delta=62.2^\circ$, and we show the function $\gamma$ 
in Fig.2.  Due to their arranged configuration, we have relatively large value 
 $|\gamma|\sim 0.8$ at the low frequency regime  $f\to 0$. The magnitude $|\gamma|$ decreases at $f\gsim 
 50$Hz, where  the  wavelength of gravitational waves becomes comparable or smaller than the 
 separation $D$ between the two detectors.

\begin{figure}
  \begin{center}
\epsfxsize=7.5cm
\begin{minipage}{\epsfxsize} \epsffile{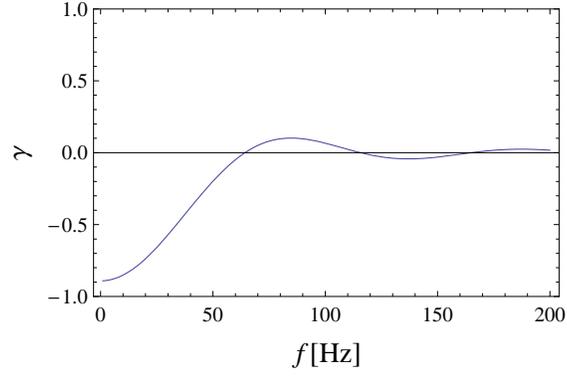}\end{minipage}
 \end{center}
  \caption{The overlap reduction function $\gamma$ for the two LIGO detectors (Hanford+Livingston).
 }
\end{figure}

For the noise spectra $P(f)=P_I(f)=P_J(f)$ of the Advanced LIGO detectors,  we use a fitting 
formula  for their broadband configuration  given in Table.1. Now, the functions 
$w_i(x)$ ($i=0,\cdots,4$) can be numerically evaluated with Eq.(\ref{wi}), 
and the shape function $w_0(x)$ is  
presented in Fig.3.  Because of our choice at $f_{\mathrm c}=25$Hz,  the curve is nearly 
flat around $x=0$.  For convenience at reproducing our numerical results below, 
we provide a fitting formula for the shape function 
\beqa
w_{0,{\mathrm{fit}}}(x)&=& 5.21169\times 10^{-8} x^{11}
+ 3.20583\times 10^{-7} x^{10} + 
1.01176\times 10^{-6} x^9 +
5.77855\times 10^{-6} x^8\nonumber\\
& & + 
0.0000303052
   x^7+0.000177148 x^6+0.000517451
   x^5+0.00412307 x^4\nonumber\\
& &+0.00506169
   x^3+0.0850738 x^2+0.0458547 x+1.00000
\eeqa
valid  in the range $x\in[-2,2]$.
While we use more accurate interpolation method for the  functions 
$w_i(x)$ throughout this paper, even the fourth derivative $\p_x^4w_{0,{\mathrm{fit}}}(x)$ 
  well approximates the accurate result $w_{4}(x)$ with error less than 0.5\% in the range
  $x\in[-2,2]$.

\begin{table}
\begin{tabular}{l|cclcl}
\hline
frequency regime [Hz] & noise spectrum $[{\rm Hz^{-1}}]$  \\
\hline
 $10\le f \le 240$ &  $10^{-44} \lmk {f }/{\rm 10 Hz}\rmk^{-4}+10^{-47.25 }
\lmk {f }/{\rm 100 Hz} \rmk^{-1.7}$  \\
\hline
$240\le f \le 3000$  & $10^{-46} \lmk {f }/{\rm 1000 Hz} \rmk^{3}  $  \\
\hline
 otherwise & $\infty$ \\
\hline
\end{tabular}
\caption{A fitting formula for the noise spectrum of the Advanced LIGO detectors (broadband 
configuration) given in \cite{Seto:2008sr}.  }
\end{table}

\begin{figure}
  \begin{center}
\epsfxsize=7.5cm
\begin{minipage}{\epsfxsize} \epsffile{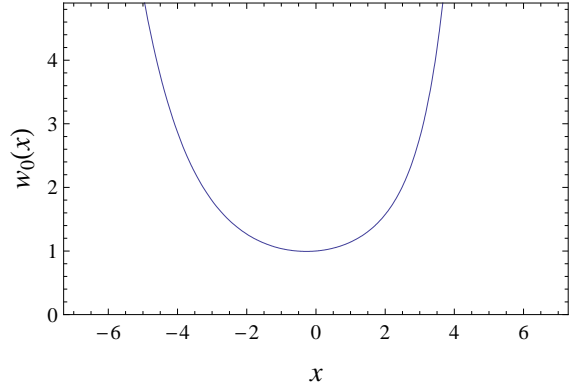}\end{minipage}
 \end{center}
  \caption{The shape function $w_0(x)$ for the correlation analysis with the 
two Advanced
LIGO detectors.
 }
\end{figure}

\begin{figure}
  \begin{center}
\epsfxsize=7.5cm
\begin{minipage}{\epsfxsize} \epsffile{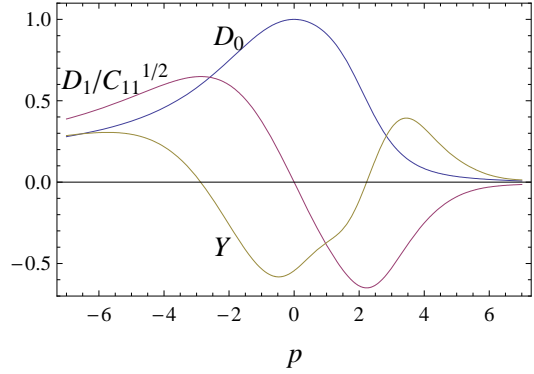}\end{minipage}
 \end{center}
  \caption{Correlation of two unit vectors $D_0=\{\hk(p),\hk(0) \}$, the ratio 
$D_1/\sqrt{C_{11}}$ and the 
inner product $Y=\{\hk(p),\hk_{\mathrm {oth}}\}$ related to the peak/valley abundances.
 }
\end{figure}

We also evaluate the factor $B$ defined in Eq.(\ref{facb}) and obtain the scaling formula 
\beq
\rho=1.536 \lmk\frac{\omega_{\mathrm{GW}}}{10^{-9}}  \rmk \lmk\frac{T_{\mathrm{obs}}}{10^{8}{\rm 
sec}}  \rmk^{1/2} [w_0(2p)]^{1/2}
\eeq
for the two Advanced LIGO detectors.  This  equation relates the amplitude $\omega_{\mathrm{GW}}$ 
of the spectrum with the SNR $\rho$.

\subsection{correlation functions}
From the numerical results $w_i(x)$ ($i=0,\cdots,4$), we can calculate the 
products $C_{ij}$ and $D_i$ using the expressions in \S IV.
In this subsection, we evaluate various correlation functions that  appear in our 
analytical expressions for the local peaks/valleys densities presented in \S III. Hereafter 
we assume that the  true GW  
background has a flat spectrum  (namely, $p_{\mathrm{t}}=0$). 

As discussed in \S II.B, the inner product $\cM_I(p)$ between the data 
$\mu=\rho_{\mathrm{t}}\hk(0)+\nu$ and the normalized template $\hk(p)$ 
is given by the mean value $\rho_{\mathrm{t}} D_0(p)$ and  the noise part $N_0(p)$ as
\beq
\cM_I=\{\mu,\hk(p)\}=\rho_{\mathrm{t}} D_0(p)+N_0(p)
\eeq
where we explicitly show the dependence on the spectral index $p$ as
\beq
D_0(p)=\{\hk(0),\hk(p)\},~~~N_0(p)=\{\nu,\hk(p)\}.
\eeq

The mean value $D_0(p)$ is identical to the correlation between the true unit vector 
$\hk(0)$ and the trial template $\hk(p)$.  As shown in Fig.4, this function takes the 
maximum value $D_0=1$ at $p=p_{\mathrm{t}}=0$, and approaches 0 at larger $|p|$.  In the 
same figure, we also show  
the quantities $D_1/\sqrt{C_{11}}$ and $Y$.  These directly appear in the 
expressions for the peak/valley densities $\sigma_{\mathrm{pk}}(p)$ and $\sigma_{\mathrm{vl}}(p)$ 
(see Eqs.(\ref{2pka}) and (\ref{2vla})), and also approach  0 at large $|p|$.
The parameter $Y$ 
characterizes the relative abundances of the local peaks and valleys through the 
functions $F_{\mathrm{pk}}$ and $F_{\mathrm{vl}}$, and they 
take similar densities at $Y\sim 0$.

{ Around the true value $p=0$, we have $Y<0$. At $\rho_{\mathrm t}\to \infty$, we need a high-$\sigma$ noise $N_{\mathrm{oth}}>-\rho_{\mathrm t}Y>0$ to make a valley 
by inverting the sign 
of the second derivative of the product $\cM_I(p)$ against the background level 
$\rho_{\mathrm t}Y$ (see Eq.(\ref{fepk}) for a related expression).  This results in a significant reduction of the valley 
density $\sigma_{\mathrm{vl}}(p)$ around $p=0$, compared with the peak density $\sigma_{\mathrm{pk}}(p)$, 
as 
shown in the next subsection. }

In Fig.4, we took the plot range up to $|p|\sim 7$ where the GW backgrounds become
extremely blue or red, and, at these ends,   it would be 
unreasonable to assume a 
single power-law spectrum $\Omega_{\mathrm{GW}}(f)$ in the whole LIGO band. But results 
in these regime 
would be instructive to see qualitatively how the correlation between the data 
and the templates affects the abundances of the local peaks and valleys.

The variance of the noise $N_0(p)$ becomes unity $\lla N_0(p)^2 \rra=1 $, and 
its correlation $\lla N_0(p_1) N_0(p_2) \rra=\{ \hk(p_1),  \hk(p_2)\}$ at 
different points $p_1$ and $p_2$ is 
presented in Fig.5. The cross-section view at $p_1=0$ is identical to the 
function $D_0(p)$ shown in Fig.4.

\begin{figure}
  \begin{center}
\epsfxsize=6.5cm
\begin{minipage}{\epsfxsize} \epsffile{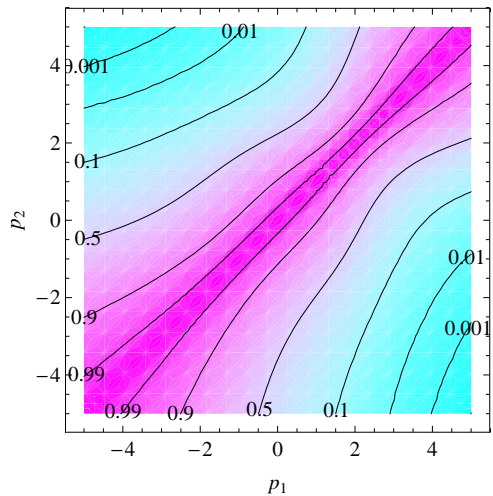}\end{minipage}
 \end{center}
\caption{Correlation of noises $\lla N_0(p_1)N_0(p_2) \rra$ 
at two points $p=p_1$ and $p=p_2$.  We have the relation $\lla N_0(p_1) N_0(p_2)\rra=\{\hk(p_1),\hk(p_2)\}$.
 }
\end{figure}

\subsection{ densities of local peaks/valleys}

Now we evaluate the statistical formulae for the local peaks/valleys derived in 
\S III.A. Hereafter, we use the expressions 
(\ref{2pka}),(\ref{2vla}),(\ref{3pka}) and (\ref{3vla}) associated with the 
local peaks/valleys of the function $\cM_I(p)$ (not Eqs.(\ref{re1}) and 
(\ref{re2}) defined for $|\cM_I(p)|$).

In Fig.6, we plot the local peak density $\sigma_{\mathrm{pk}}(p)$ for the intrinsic signal 
strengths $\rho_{\mathrm{t}}=0,1,2,4$ and 8. For larger $\rho_{\mathrm{t}}$, the peak density shows 
stronger concentration around the true value $p=0$.  
We can also observe increment of the density $\sigma_{\mathrm{pk}}$ around $p\gsim 6$ 
where the ratio $|D_1/\sqrt{C_{11}}|$ decreases again (see Fig.4) in 
eq.(\ref{2pka}) with $D_1(p)=\p_p D_0(p)\simeq 0$ reflecting
$D_0(p)\simeq0$. Notice that the exponential term of the
r.h.s of Eq.(\ref{2pka}) takes the maximum value at
$|D_1/\sqrt{C_{11}}|=0$ and the function $F_{\mathrm{pk}} ( - \rho_{\mathrm{t}} Y )$ becomes
a constant at $Y=0$.
This increment of  $\sigma_{\mathrm{pk}}$ is mainly caused by the noise, as examined later.

We show the density of the local valleys in Fig.6. This  function is strongly
suppressed by the Gaussian-like factor $F_{\mathrm{vl}} (- \rho_{\mathrm{t}} Y)$ (see Eq.(\ref{fvl})) 
around the true value $p=0$.  Therefore, for signal strength $\rho_{\mathrm{t}}\gg 1$, it is very 
unlikely that there exist multiple local peaks around $p=0$, 
since we must have a valley between  two peaks.  We can further expect that the peak 
identified around $p=0$ is likely to be the global one that we want to identify at data analysis.

In order to support this from  the viewpoint of the total numbers of local 
peaks/valleys around $p\sim 0$, we define the integrals (with the integration range selected 
somewhat arbitrarily) as
\beq
U_{\mathrm{pk}}\equiv \int_{-3}^3 \sigma_{\mathrm{pk}}(p)dp,~~~U_{\mathrm{vl}}\equiv \int_{-3}^3 \sigma_{\mathrm{vl}}(p)dp.
\eeq
The results are shown in Fig.7.  We have $U_{\mathrm{pk}}<1$ at $\rho_{\mathrm{t}}<3$, but  
$U_{\mathrm{pk}}>1$ at $\rho_{\mathrm{t}}>3$. The result $U_{\mathrm{pk}}>1$
suggests that, in principle,  the spurious peaks due to the noise appear in the range
$-3<p<3$.  But  the expected number of the peaks in the 
range  becomes nearly unity $|U_{\mathrm{pk}}-1|\le 10^{-5}$ at 
$\rho_{\mathrm{t}}>8$. { The asymptotic slope is steeper than the weak bound $O(\rho_{\mathrm{t}}^{-2})$ given in Eq.(\ref{mean0}). }

\begin{figure}
  \begin{center}
\epsfxsize=7.5cm
\begin{minipage}{\epsfxsize} \epsffile{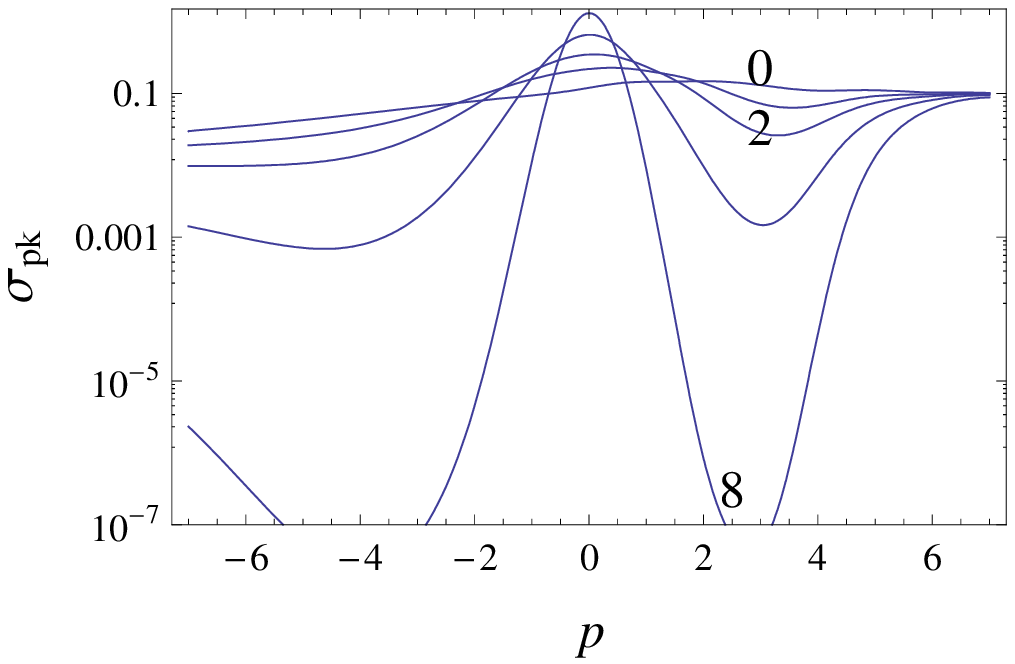}\end{minipage}
\epsfxsize=7.5cm
\begin{minipage}{\epsfxsize} \epsffile{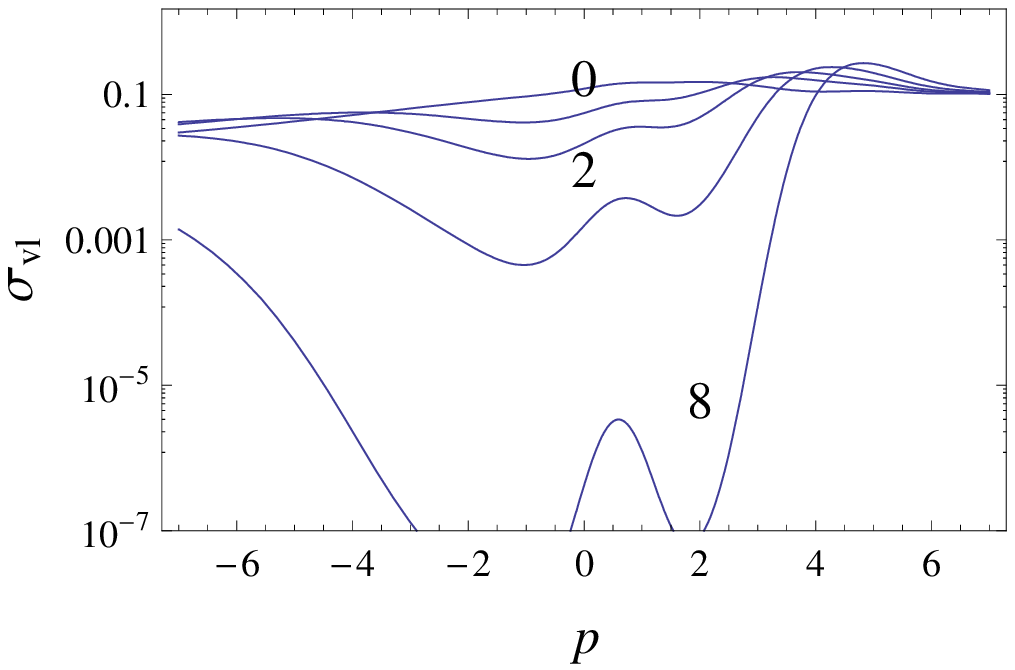}\end{minipage}
 \end{center}
  \caption{Density distribution of the local peaks $\sigma_{\mathrm{pk}}(p)$ and valleys  $\sigma_{\mathrm{vl}}(p)$  for the true 
spectral index $p_{\mathrm{t}}=0$. We plot five curves for the intrinsic signal strengths 
$\rho_{\mathrm{t}}=0,1,2,4$ and 8.
 }
\end{figure}

\begin{figure}
  \begin{center}
\epsfxsize=7.5cm
\begin{minipage}{\epsfxsize} \epsffile{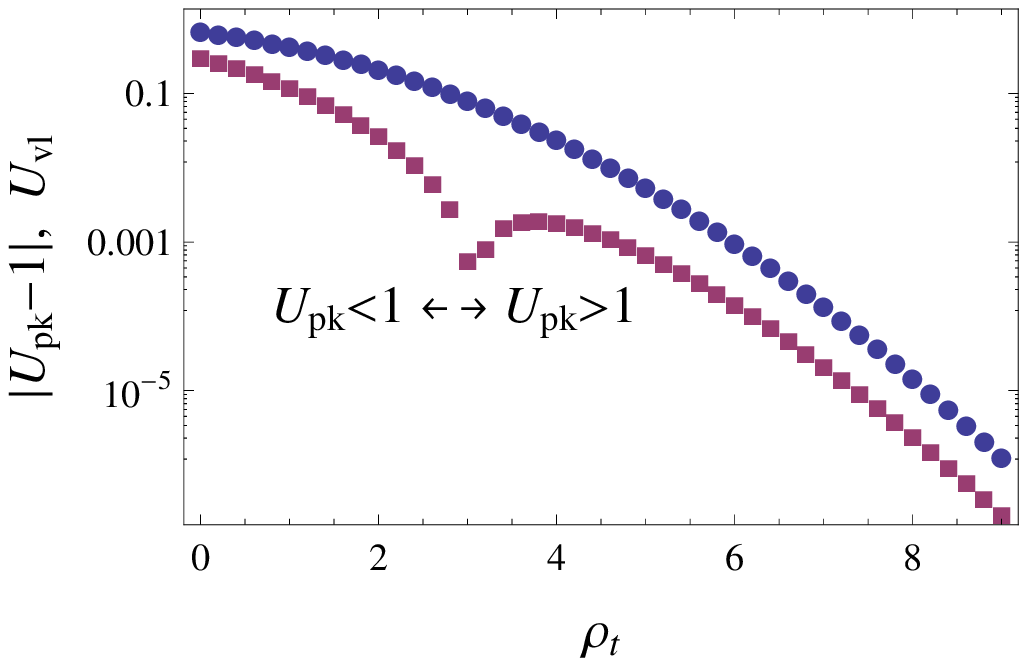}\end{minipage}
 \end{center}
  \caption{Profiles of the integrals $U_{\mathrm{pk}}$ (squares) and $U_{\mathrm{vl}}$ (circles). We have $U_{\mathrm{pk}}<1$ 
at $\rho_{\mathrm{t}}\lsim3$., and $U_{\mathrm{pk}}>1$ at $\rho_{\mathrm{t}}\gsim3$. 
 }
\end{figure}

\begin{figure}[!h]
  \begin{center}
\epsfxsize=7.5cm
\begin{minipage}{\epsfxsize} \epsffile{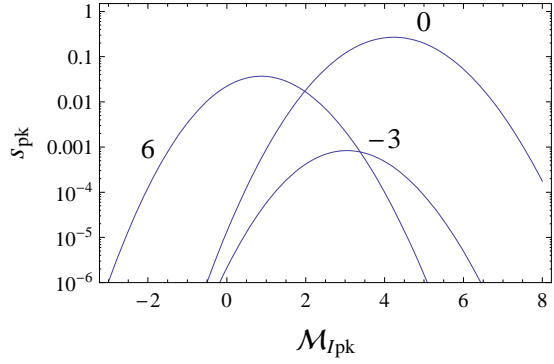}\end{minipage}
 \end{center}
  \caption{Distribution $s_{\mathrm{pk}}(\cM_{I \mathrm{pk}},p)$ of the peaks height $\cM_{I \mathrm{pk}}$  identified at the  
spectral indexes $p=-3,0$ and 6.  The intrinsic signal strength is $\rho_{\mathrm{t}}=4$. 
We have the identity $\int_{-\infty}^\infty d\cM_{I \mathrm{pk}}s_{\mathrm{pk}}(\cM_i,p)=\sigma_{\mathrm{pk}}(p)$ for the area of each 
curve.
 }
\end{figure}

As commented earlier, 
the values $\cM_{I \mathrm{pk}}$  at the local peaks  themselves are the primary indicator 
at actual  data 
analysis. The peak with the maximum value $\cM_{I \mathrm{pk}}$ should be selected 
among multiple local peaks.
The peaks  existing around 
the true value $p=0$ would have relatively large values due to the underling 
correlation $D_0(p)$ before the noise $N_0(p)$ is added.
 We thus examine the distribution of the height  $\cM_{I \mathrm{pk}}$ of  the 
local peaks 
identified   at a given spectral index $p$.   In Fig.8 we plot 
examples of the profile $s_{\mathrm{pk}}(\cM_{I \mathrm{pk}},p)$ (see Eq.(\ref{3spk})) for $\rho_{\mathrm{t}}=4$ at the specific 
spectral indexes $p=-3$,0 and 6. 
Even if two local peaks are simultaneously identified {\it e.g.} at $p\sim0$ and 
$p\sim 6$, the desired one $p\sim 0$ would be appropriately selected on 
the ground of the magnitude  $\cM_{I
\mathrm{pk}}$, as far as $\rho_\mathrm{t} \gtrsim 3$ so that peaks for
true and spurious values of $p$ are likely to be distinguished (see Eq.~(\ref{3pka})).


\begin{figure}[t]
\begin{center}
\includegraphics[width=5.cm,clip]{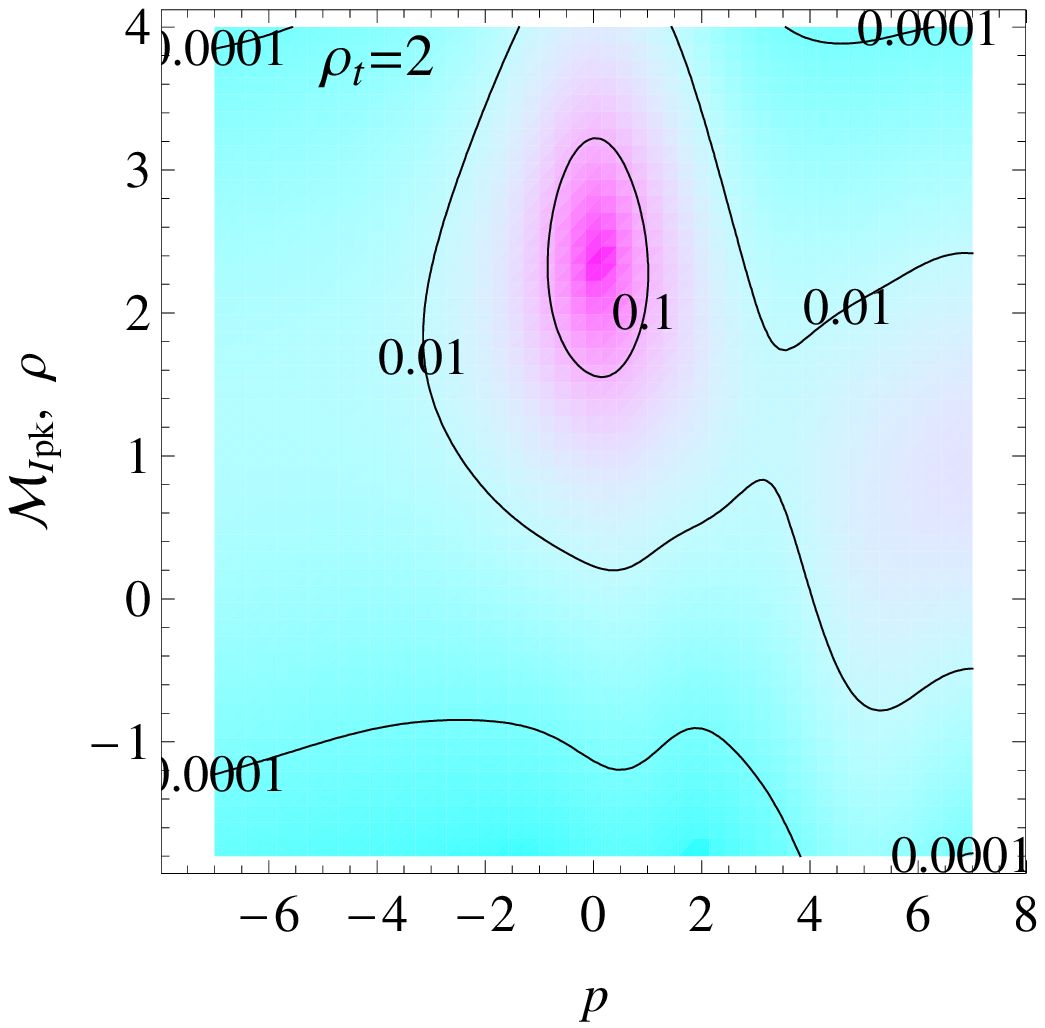}
\hspace*{0.5cm}
\includegraphics[width=5.cm,clip]{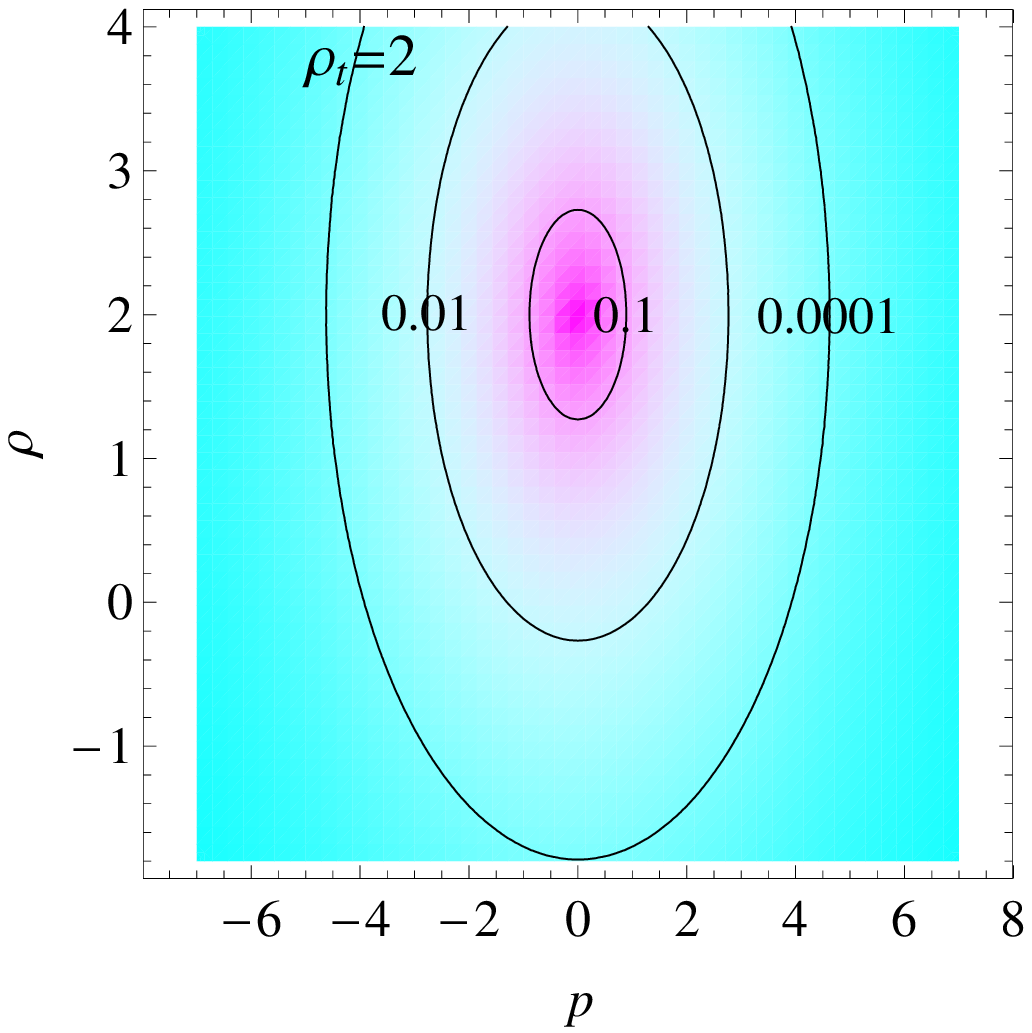}
\hspace*{0.5cm}
\includegraphics[width=5.cm,clip]{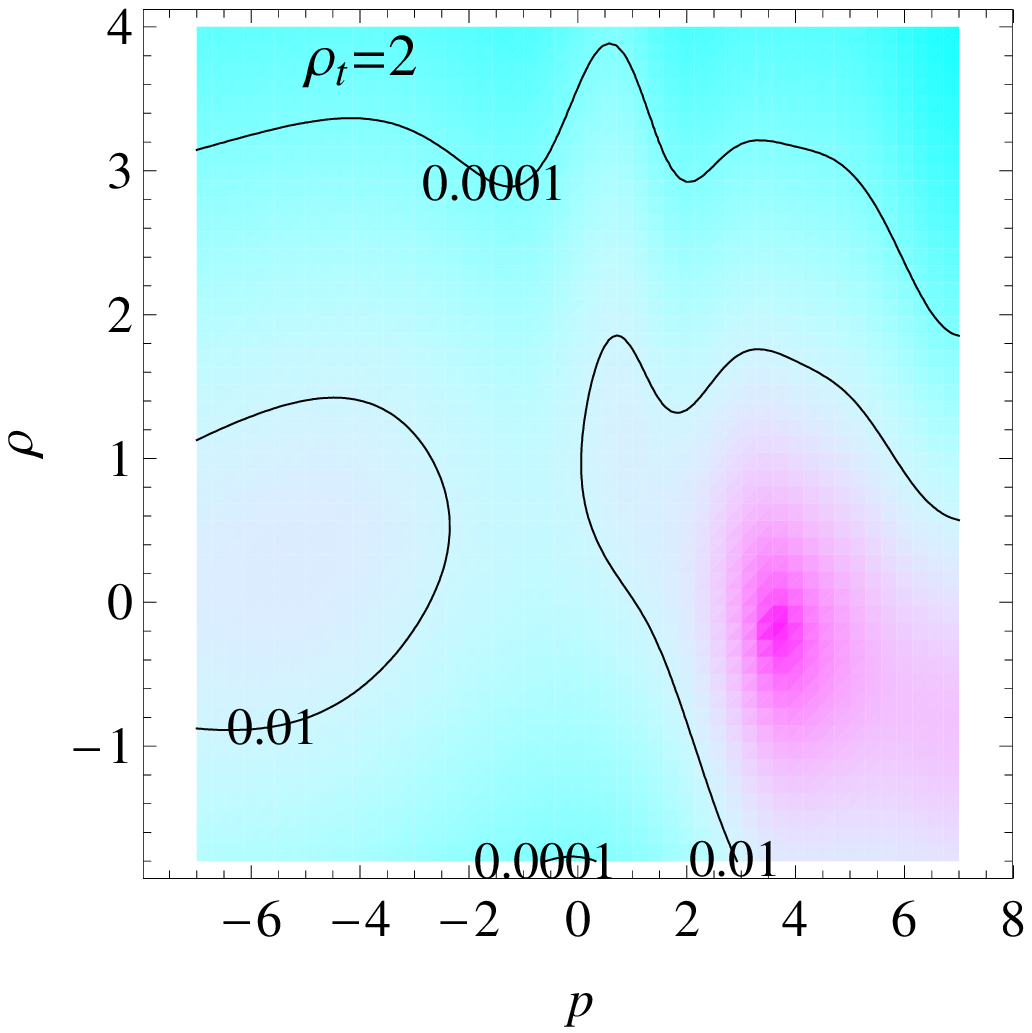}
\end{center}

\begin{center}
\includegraphics[width=5.cm,clip]{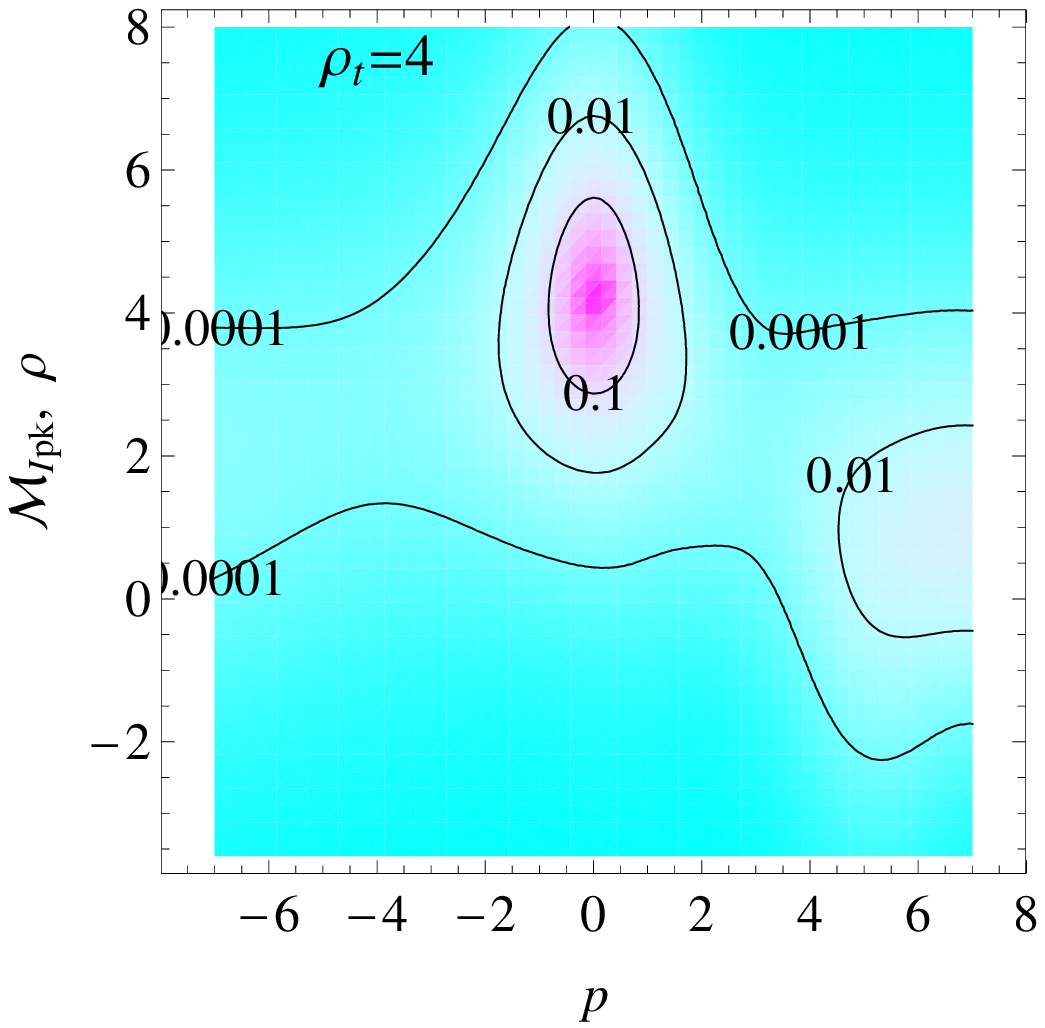}
\hspace*{0.5cm}
\includegraphics[width=5.cm,clip]{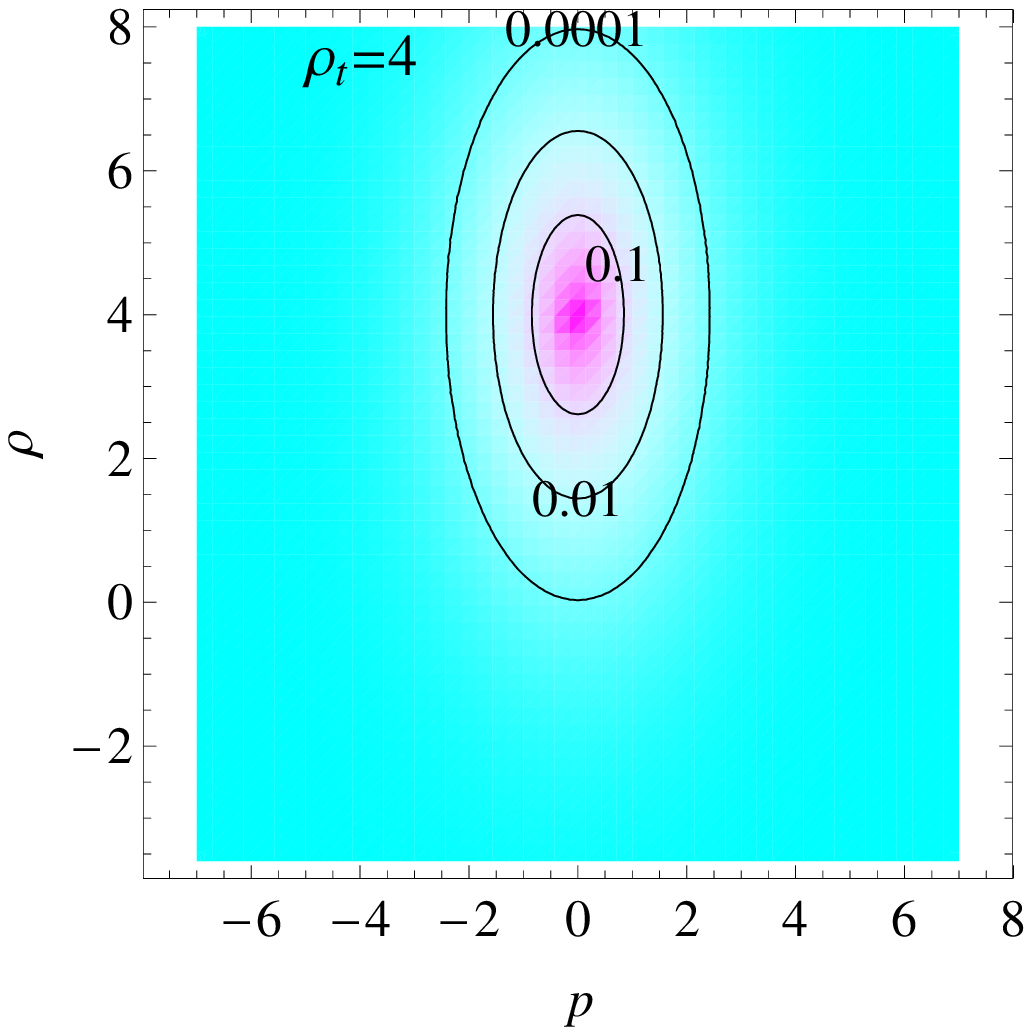}
\hspace*{0.5cm}
\includegraphics[width=5.cm,clip]{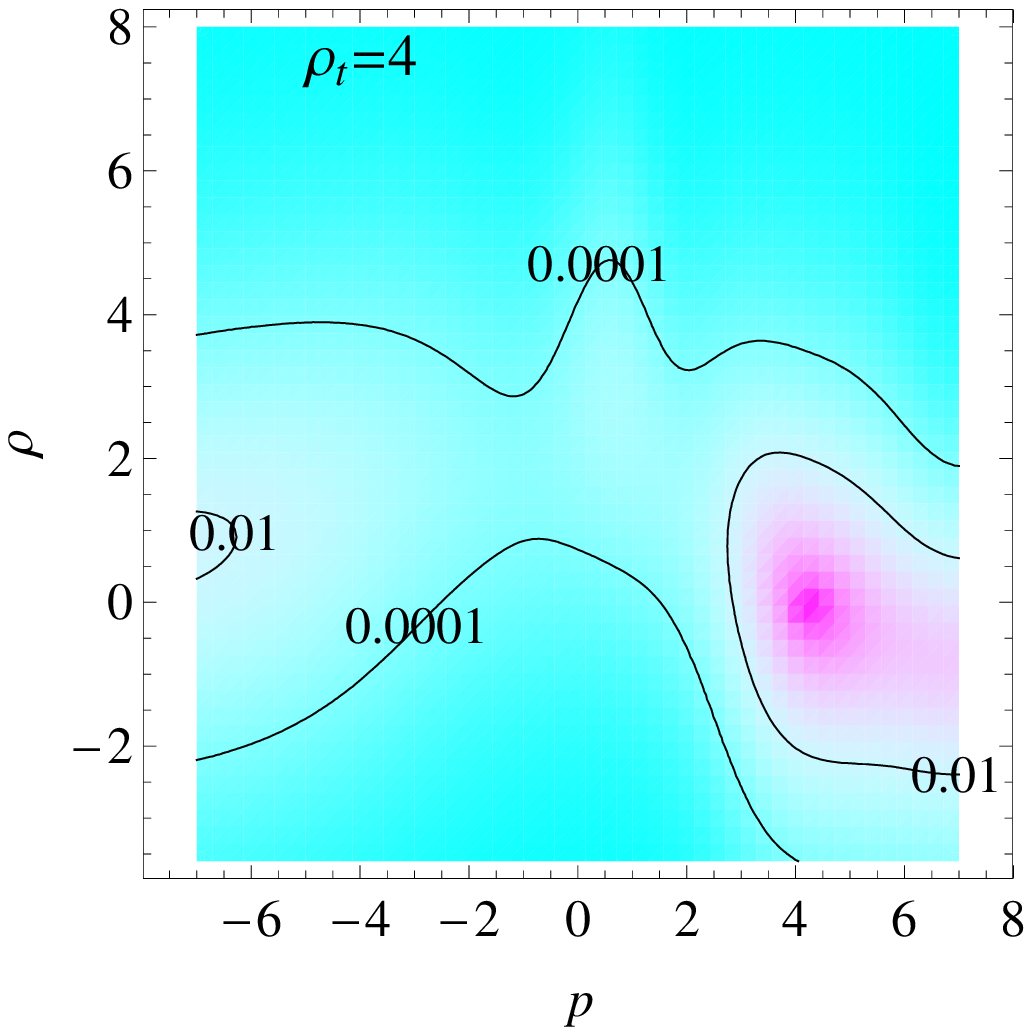}
\end{center}

\begin{center}
\includegraphics[width=5.cm,clip]{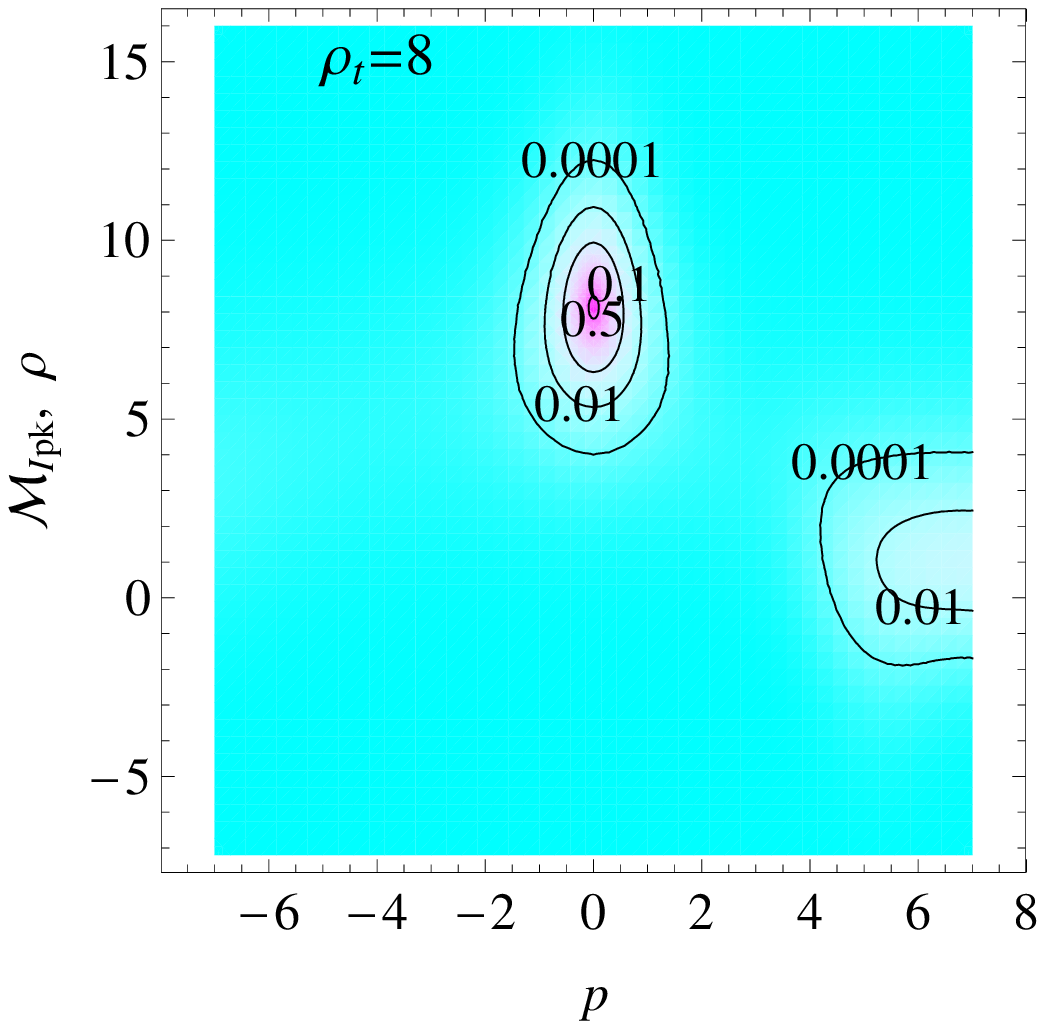}
\hspace*{0.5cm}
\includegraphics[width=5.cm,clip]{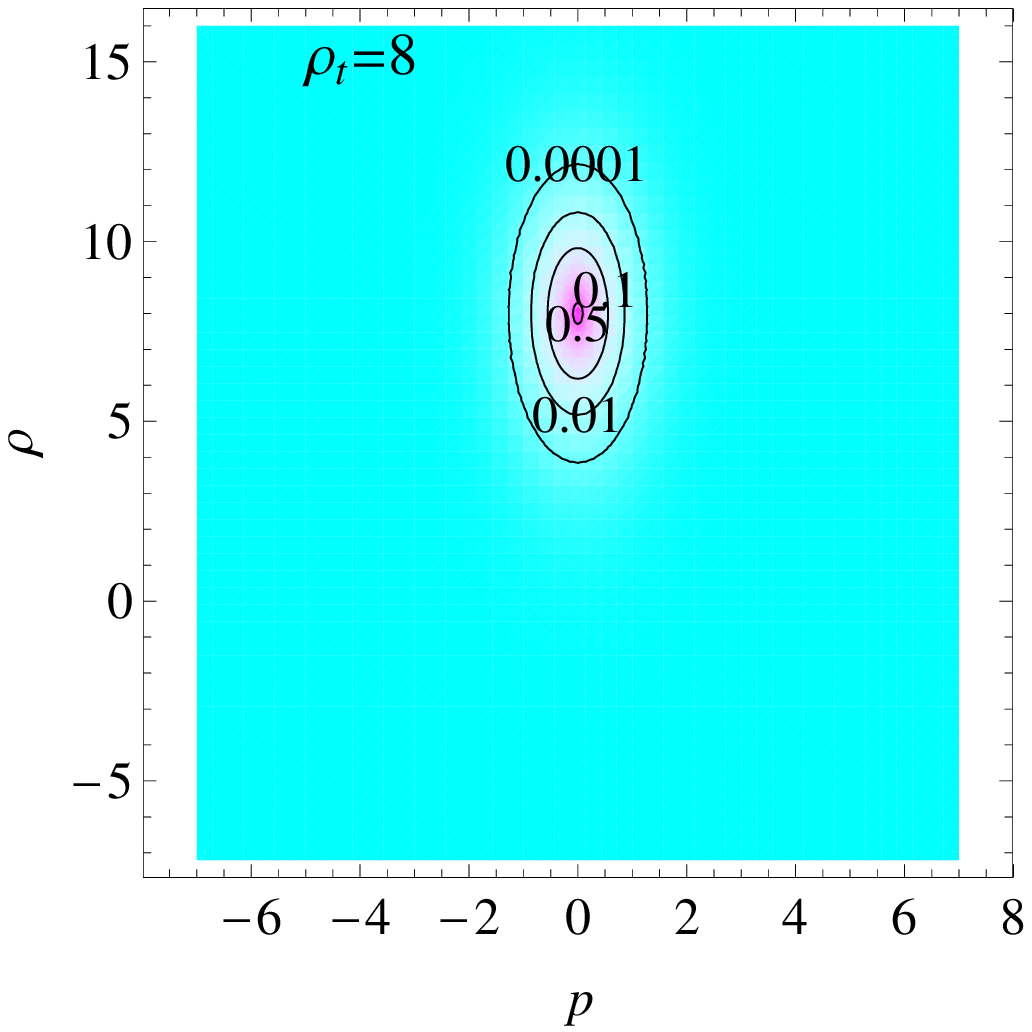}
\hspace*{0.5cm}
\includegraphics[width=5.cm,clip]{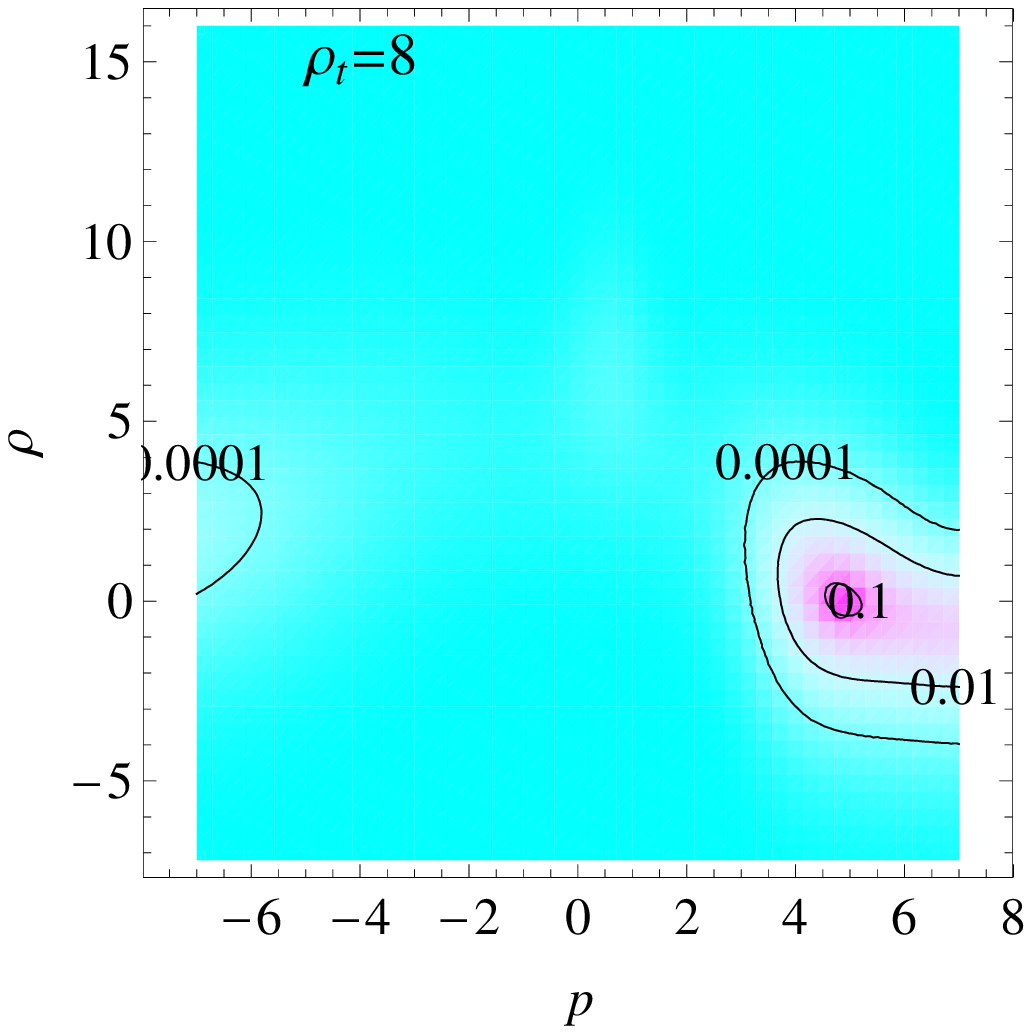}
\end{center}

\vspace*{-0.2cm}

\caption{The two dimensional density distribution 
$\sigma_{\mathrm{pk}}(\rho,p)=\sigma_{\mathrm{pk}}(\cM_{I \mathrm{pk}}=\rho,p)$ of local peaks (left), the Fisher matrix 
prediction $\sigma_{\mathrm{pk}}(p)$ (middle) and the density of saddle points 
$\sigma_{\mathrm{vl}}(\rho,p)$ (right). The true spectral index is $p=p_{\mathrm{t}}=0$ and 
the  intrinsic signal strength is at  
$\rho_{\mathrm{t}}=2$ (top row), 4 (middle row) and 8 (bottom row).  We show the isodensity contours for 
$\sigma_{\mathrm{pk}}=0.5, 0.1,0.01$ and 0.0001.}
\label{t1}
\end{figure}

\begin{figure}
  \begin{center}
\epsfxsize=7.5cm
\begin{minipage}{\epsfxsize} \epsffile{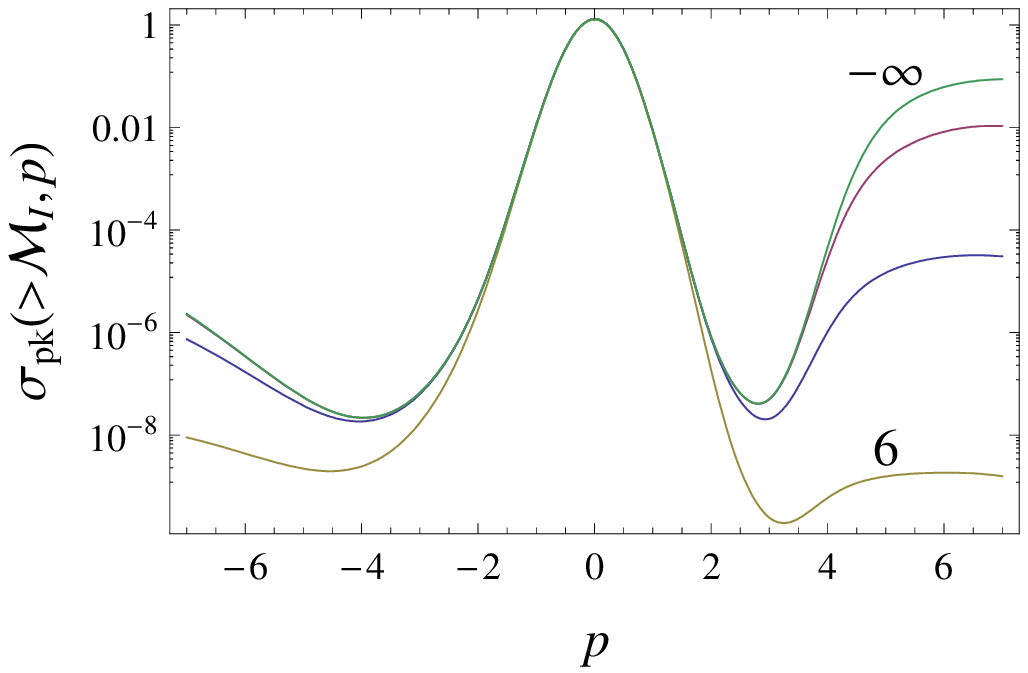}\end{minipage}
 \end{center}
  \caption{ The densities of the local peaks $\sigma_{\mathrm{pk}}(>\cM_{I \mathrm{pk}},p)$ above given 
thresholds $\cM_{I \mathrm{pk}}=-\infty,2,4$ and 6.
 }
\end{figure}

In the left panels of Fig.9, we show the two dimensional contour plots for 
the function $s_{\mathrm{pk}}(\cM_{I \mathrm{pk}},p)$ (identical to $\sigma_{\mathrm{pk}}(\rho=\cM_{I \mathrm{pk}},p)$ 
and discussed later).
We can observe high density region around $(\cM_{I \mathrm{pk}},p)=(\rho_{\mathrm{t}},p_{\mathrm{t}})$, and an
additional increment around $\cM_{I \mathrm{pk}}\sim 0$ and  $p\gsim 6$.  The latter is due to 
the local peaks mainly caused by 
the noise. For $\rho_{\mathrm{t}}=8$, these two are well separated and the latter 
 would not survive at the data analysis where we check the magnitude $\cM_{I \mathrm{pk}}$.

In order to show this explicitly, in Fig.10, we plot the local peak density 
$\sigma_{\mathrm{pk}}(>\cM_{I \mathrm{pk}},p)$ above given threshold $\cM_{I \mathrm{pk}}$ (see Eq.(\ref{pkth}) for its
definition). With the identity  
$\sigma_{\mathrm{pk}}(>-\infty,p)=\sigma_{\mathrm{pk}}(p)$, the uppermost curve  is the same as 
 the unconstrained one $\sigma_{\mathrm{pk}}(p)$ given in Fig.8 for $\rho_{\mathrm{t}}=8$. The 
abundance of the local peaks around the true value $p=0$ is nearly the same for the four 
curves.  But  the local peaks at $p\gsim 6$ have the 
typical value $\cM_{I \mathrm{pk}}\sim 
0$ and most of them are removed for the suitable threshold $\cM_{I \mathrm{pk}}=4$.

In Fig.9, the densities $\sigma_{\mathrm{pk}}(\cM_{I \mathrm{pk}},p)$ take their maximum
values at the points with $\cM_{I \mathrm{pk}}>\rho_{\mathrm{t}}$.  This can be directly
confirmed by putting $p=0$ in Eq.(\ref{3pka}) with $D_0=1$ and $D_1=0$,
and using the monotonic shape of the function
$F_{\mathrm{pk}}$. { This overestimation is closely related to the bias (see Eq.(\ref{meana})) of the 
amplitude parameter $\rho$ discussed in the next subsection.}

\subsection{Fisher matrix predictions}

Here we compare our local peak density $\sigma_{\mathrm{pk}}(p)$ with the Fisher matrix 
prediction $\sigma_{\mathrm{fisher}}(p)$ defined in Eq.(\ref{fish}). The examples are shown in 
Fig.11. At $\rho_{\mathrm{t}}\gsim 4$, the Gaussian-like profiles around the true value 
$p=0$ are similar for the two curves, and this indicates  that  
the simple Fisher matrix prediction becomes a reasonable approximation in this 
regime.

In Fig.12, we show the difference  
$\sigma_{\mathrm{pk}}-\sigma_{\mathrm{fisher}}$ between two expressions. Since the local peak 
density  $\sigma_{\mathrm{pk}}$  works as  an upper limit for the probability
distribution function of the global peaks, the Fisher matrix predictions over 
estimates the probability distribution function at the spectral indexes $p$  
with $\sigma_{\mathrm{pk}}-\sigma_{\mathrm{fisher}}<0$.

In Fig.13, we plot the function $\sigma_{\mathrm{pk}}-\sigma_{\mathrm{fisher}}$ now using the 
rescaled parameter $x\equiv (\rho_{\mathrm{t}} \Delta p)$ introduced in \S III.B.  As 
expected from the analytical evaluation, the difference  
$\sigma_{\mathrm{pk}}-\sigma_{\mathrm{fisher}}$ approaches the leading order correction 
$\eta(x)$ which is an odd function and characterized by  two parameters 
$C_{11\mathrm{t}}=0.168$ and $C_{21\mathrm{t}}=0.007154$ in the present case.

Next we calculate the mean and variance  of the local peak distribution. To this 
end, we take the parameter range $p\in[-3,3]$ and renormalize the peak density 
as $\sigma_{\mathrm{pk}}/U_{\mathrm{pk}}$ to regard it as a probability distribution function. We 
then evaluate the integrals
\beq
\lla \Delta p\rra=\frac1{U_{\mathrm{pk}}}\int_{-3}^3\sigma_{\mathrm{pk}}(\Delta 
p) d\Delta p,~~\lla (\Delta 
p)^2\rra=\frac1{U_{\mathrm{pk}}}\int_{-3}^3\sigma_{\mathrm{pk}}(\Delta p)^2 
d\Delta p. \label{meana2}
\eeq
The results  at various $\rho_{\mathrm{t}}$ are shown in Fig.14, along with the 
leading order contributions ($\propto \rho_{\mathrm{t}}^{-2}$) given in the left hand sides 
of Eqs.(\ref{mean}) and (\ref{vari}).
 Note that the mean value $\lla \Delta p\rra$
changes its sign around $\rho_{\mathrm{t}}\sim4$ (from $+$ to $-$).  The analytical curves 
show  good agreements with the numerical ones at $\rho_{\mathrm{t}}\gsim 5$.
{ Similarly, we evaluate the integral 
\beq
\lla \Delta \rho 
\rra=\frac1{U_{\mathrm{pk}}}\int_{-3}^3\sigma_{\mathrm{pk}}[{\bar 
\rho}_{\mathrm{pk}}-\rho_{\mathrm t}] d\Delta p \label{meana3}
\eeq
for the mean bias for the amplitude parameter $\rho$. In Fig.15, we plot the 
results with the asymptotic expression $1/(2\rho_{\mathrm t})$ given in Eq.(\ref{meana}).  These two also 
show a good agreement.}

\begin{figure}[h]
\begin{center}
\includegraphics[width=5.cm,clip]{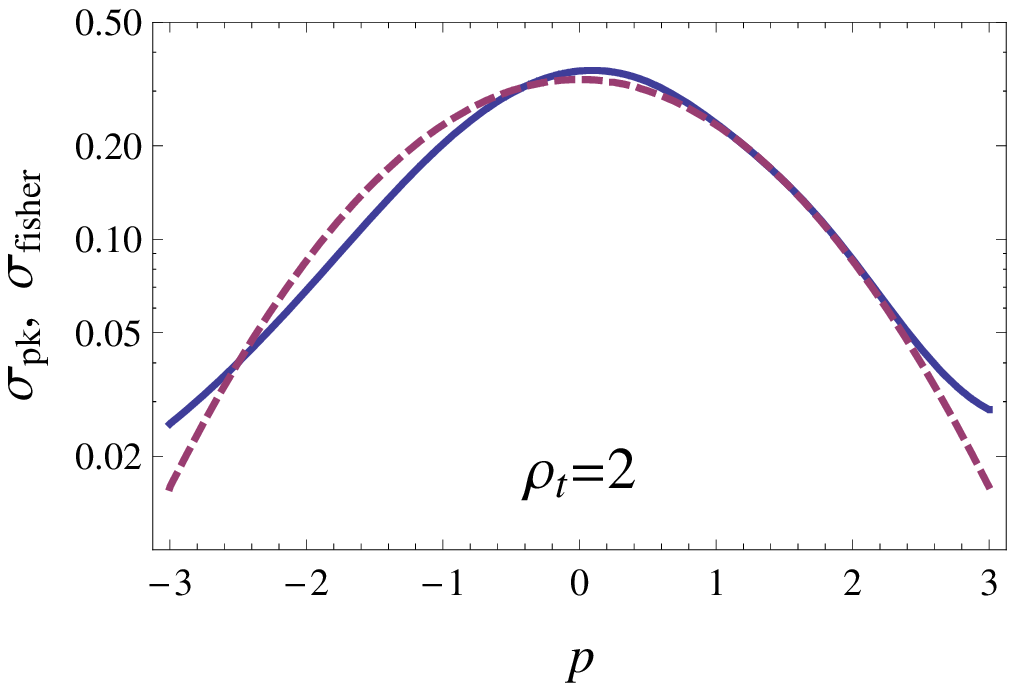}
\hspace*{0.5cm}
\includegraphics[width=5.cm,clip]{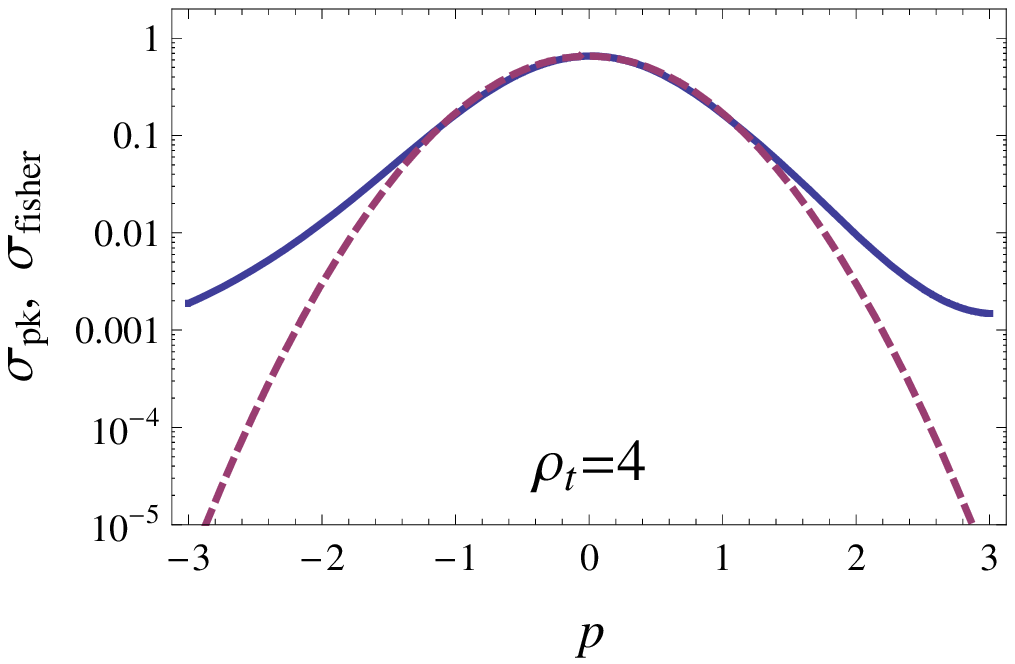}
\hspace*{0.5cm}
\includegraphics[width=5.cm,clip]{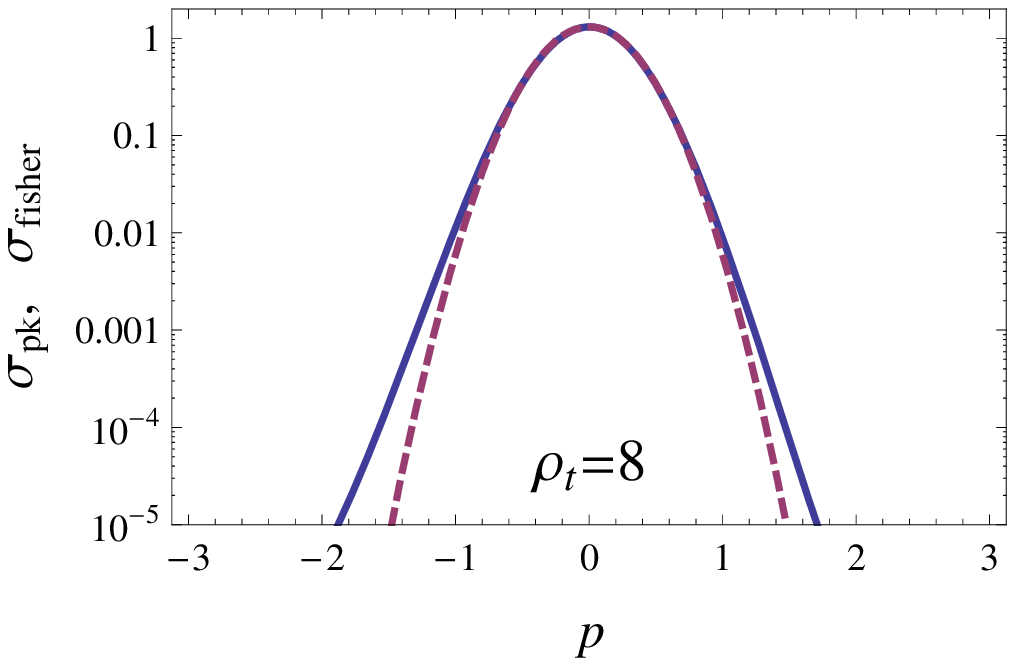}
\end{center}

\vspace*{-0.2cm}

\caption{Comparison of the local peak density $\sigma_{\mathrm{pk}}(p)$ (solid curves) with the Fisher 
matrix prediction (dashed curves). }
\label{t1}
\end{figure}

\begin{figure}
  \begin{center}
\epsfxsize=7.5cm
\begin{minipage}{\epsfxsize} \epsffile{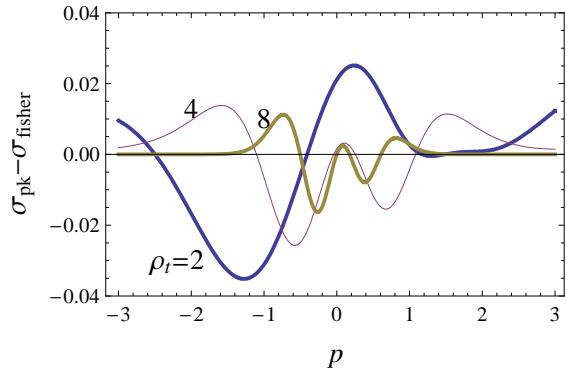}\end{minipage}
 \end{center}
  \caption{Difference between the local peak density $\sigma_{\mathrm{pk}}(p)$ and the 
Fisher matrix prediction $\sigma_{\mathrm{fisher}}(p)$ for the intrinsic signal 
strengths at $\rho_{\mathrm{t}}=2,4$ and 8.
 }
\end{figure}

\begin{figure}
  \begin{center}
\epsfxsize=7.5cm
\begin{minipage}{\epsfxsize} \epsffile{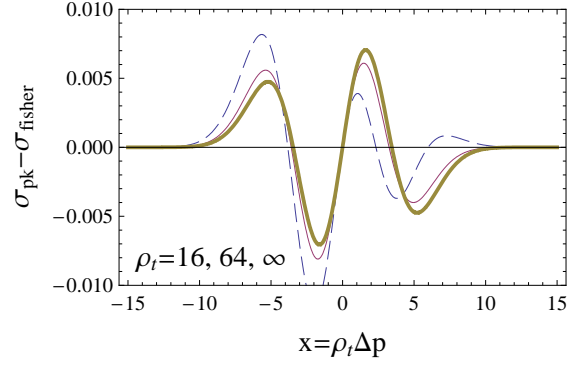}\end{minipage}
 \end{center}
  \caption{Difference  $\sigma_{\mathrm{pk}}(p)-\sigma_{\mathrm{fisher}}(p)$ shown with the 
rescaled variable $x\equiv \rho_{\mathrm{t}} \Delta p$. The dashed curve is for $\rho_{\mathrm{t}}=16$ 
and the thin solid one for $\rho_{\mathrm{t}}=32$. The thick solid one is the asymptotic 
limit $\eta(x)$ given in Eq.(\ref{eta}). 
 }
\end{figure}

\begin{figure}
  \begin{center}
\epsfxsize=7.5cm
\begin{minipage}{\epsfxsize} \epsffile{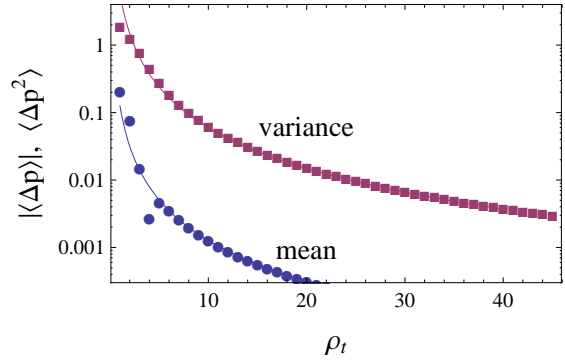}\end{minipage}
 \end{center}
  \caption{Absolute values for the variance and mean of the local peak density 
(see Eq.(\ref{meana2})). The solid curves are the 
leading order terms $\propto \rho_{\mathrm{t}}^{-2}$ (Eqs.(\ref{mean}) and (\ref{vari})). The mean value changes its sign from 
$+$ to $-$ around $\rho_{\mathrm{t}}=4$.
 }
\end{figure}

\begin{figure}
  \begin{center}
\epsfxsize=7.5cm
\begin{minipage}{\epsfxsize} \epsffile{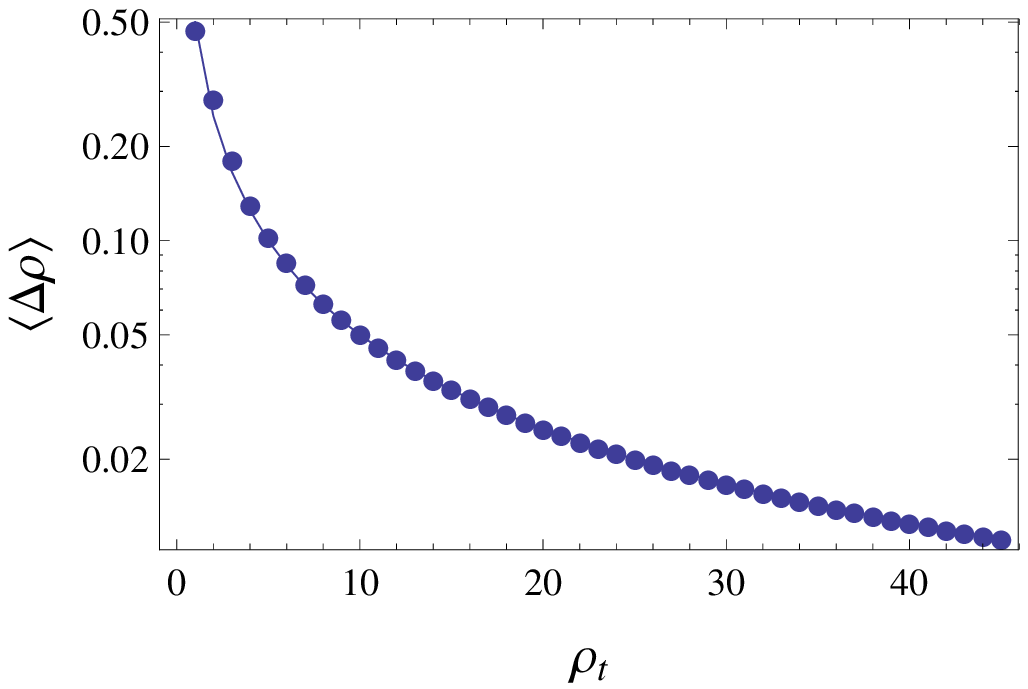}\end{minipage}
 \end{center}
  \caption{ The positive bias $\lla \rho \rra$ for the amplitude parameter $\rho$ estimated by 
the maximum likelihood analysis. The circles are obtained with Eq.(\ref{meana3}) 
and the solid curve is their asymptotic form $1/2\rho_{\mathrm t}$ (see Eq.(\ref{meana})).
 }
\end{figure}

With the identity $\sigma_{\mathrm{pk}}(\rho,p)=s_{\mathrm{pk}}(\cM_{I \mathrm{pk}}=\rho,p)$ mentioned at 
the end of \S III.A, the left panels in 
Fig.9 can be used to discuss the local peak distribution $\sigma_{\mathrm{pk}}(\rho,p)$ in 
the two dimensional parameter space $(\rho,p)$.  The overall behaviours of these 
figures were already described, and we do not repeat them again. 
But, in the  middle column of Fig.9, we provide the Fisher matrix predictions. 
As for the one-dimensional case shown in Fig.11, the Fisher matrix predictions 
become good approximations for larger $\rho_{\mathrm{t}}$. In the right  panels of Fig.9, 
we also show the density of saddle points $\sigma_{\mathrm{vl}}(\rho,p)$. We can observe 
increment of the density $\sigma_{\mathrm{vl}}(\rho,p)$ around the regions where the local peak density is 
enhanced by the noises (especially for $\rho_{\mathrm{t}}=8$).  Since the intrinsic 
correlation $D_0(p)$ is weak here, the preference of the sign $\p_p^2 \{\hk,\mu\}$  
is decreased and the peak/valley densities show similar patterns. In contrast, 
the saddle points are strongly suppressed around the true value $(\rho_{\mathrm{t}},p_{\mathrm{t}})$, 
as mentioned earlier.

\section{Summary}

In  this  paper, we discussed  a simplified model of data analysis where we 
estimate a single intrinsic parameter $p$ and the overall amplitude $\rho$ of a
signal that is  contaminated by Gaussian noises.  The approach behind our study was 
recently proposed by Vallisneri \cite{Vallisneri:2011ts}, and based on the fact that the local stationary 
points on the likelihood surfaces can be studied with a small number of 
independent noise components. 


In this paper, we paid special attention to the local geometric aspects of the 
likelihood surfaces, including valleys and saddle points. With our 
 analytic expressions derived owing to the simplified settings, we can see 
how the geometrical structure depend on the signal strength, the likelihood value, 
and correlation between the true and the trial templates. We expect that our 
qualitative results would provide us useful insights when dealing with more 
complicated problems of data analysis for GW astronomy (and beyond).

In the later half of this paper, we applied our formal expressions to 
correlation analysis of stochastic GW backgrounds. Considering  ubiquitously 
realized scaling behaviours of cosmological processes (and also astrophysical 
ones related to GWs), it would be reasonable to assume   a power-law 
spectrum for  
the background in the frequency regime of a GW detector and discuss accuracy of 
parameter estimation for the spectral index and the amplitude of the spectrum. 
Therefore,  the correlation analysis 
for the background can be regarded as an exemplary as well as realistic case 
for applying our formal expressions. 
At the same time, this concrete example would conversely help us to see the 
qualitative trends of the formal results.

To link the correlation analysis with the formal results, we provided useful 
expressions, including ready-to-use fitting formulae for the two LIGO detectors.
Then, we numerically evaluated the expected densities of the local 
peaks/valleys/saddle points of the likelihood surfaces.  We find that the 
abundance of the local valleys is strongly suppressed around the true parameters, 
indicating prohibition of multiple peaks there. In contrast, at the region where 
the true signal lose correlation, there appears peaks and valleys mainly caused 
by the fluctuations of the noise. These false peaks would typically have low 
likelihood significance due to the lack of the underlying signal
correlation. Therefore, they will be safely ruled out in
the actual data analysis by setting an appropriate threshold on the
value of the likelihood. At $\rho_{\mathrm{t}} \gtrsim 5$, the expansion around the
Fisher matrix prediction to the first order in $\rho_{\mathrm{t}}$ is found to
approximate the exact results to a good accuracy. 
{ We also analyzed the  biases for parameters estimated with the maximum likelihood 
method.  At $\rho_{\mathrm{t}} \gtrsim 5$, our results show good agreements with those obtained in a perturbative 
method as  second order corrections $O(\rho_{\mathrm{t}}^{-2})$ relative to the true parameters.
 }

This
work was supported by JSPS grants 20740151, 21684014, and 24540269.

\appendix

\end{document}